\documentclass[11pt]{article}
\pdfoutput=1
\usepackage{amssymb}
\usepackage{mathtools}
\usepackage{amsmath}
\usepackage{amstext}
\usepackage{graphicx,epsfig}
\usepackage{epsfig}
\usepackage{verbatim} 
\usepackage{makecell}
\usepackage{caption}
\usepackage{fancybox}
\usepackage{slashed}
\usepackage{color}
\usepackage{enumitem}	
\usepackage{subfigure}
\usepackage{bbm}
\usepackage{parskip}
\usepackage{dsfont}
\usepackage{tabu}
\usepackage[numbers,sort&compress]{natbib}

\usepackage[table]{xcolor} 

\newcommand{\Comment}[1]{{}}
\definecolor{darkblue}{rgb}{0.15,0.35,0.55}
\definecolor{reddish}{rgb}{0.65, 0.2, 0.2}
\usepackage[linktocpage=true]{hyperref}
\hypersetup{
colorlinks=true,
citecolor=darkblue,
linkcolor=reddish,
urlcolor=darkblue,
pdfauthor={},
pdftitle={},
pdfsubject={}
}

\newcommand{\be}{\begin{equation}}
\newcommand{\ee}{\end{equation}}
\newcommand{\bea}{\begin{align}}
\newcommand{\eea}{\end{align}}

\newcommand{\rd}{{\rm d}}

\newcommand{\Tr}{\, {\rm Tr} \, }

\newcommand{\nn}{ \nonumber\\}

\title{}
\author{}

\numberwithin{equation}{section}
\allowdisplaybreaks[1]

\begin{document}
%
~
\vspace{1truecm}
\renewcommand{\thefootnote}{\fnsymbol{footnote}}
\begin{center}
{\huge \bf{Heavy Fields and Gravity}}
\end{center} 

\vspace{1truecm}
\thispagestyle{empty}
\centerline{\Large Garrett Goon\footnote{\tt G.L.Goon@uva.nl} }
\vspace{.7cm}

\centerline{\it Institute of Physics, Universiteit van Amsterdam,}
\centerline{\it Science Park 904, Amsterdam, 1090 GL, The Netherlands}

\vspace{.5cm}
\begin{abstract}
\vspace{.03cm}
\noindent
We study the effects of heavy fields on 4D spacetimes with flat, de Sitter and anti-de Sitter asymptotics.  At low energies, matter generates specific, calculable higher derivative corrections to the GR action which perturbatively alter the Schwarzschild-$(A)dS$ family of solutions.    The effects of massive scalars, Dirac spinors and gauge fields are each considered.  The six-derivative operators they produce, such as  $\sim R^{3}$ terms, generate the leading corrections. The induced changes to horizon radii, Hawking temperatures and entropies are found. Modifications to the energy of large $AdS$ black holes are derived by imposing the first law. An explicit demonstration of the replica trick is provided, as it is used to derive black hole and cosmological horizon entropies. Considering entropy bounds, it's found that scalars and fermions increase the entropy one can store inside a region bounded by a sphere of fixed size, but vectors lead to a decrease, oddly.  We also demonstrate, however, that many of the corrections fall below the resolving power of the effective field theory and are therefore untrustworthy.  Defining properties of black holes, such as the horizon area and Hawking temperature, prove to be remarkably robust against higher derivative gravitational corrections.
\end{abstract}

\newpage
\tableofcontents

\renewcommand*{\thefootnote}{\arabic{footnote}}
\setcounter{footnote}{0}

\section{Heavy Fields and GR}

How do heavy fields affect gravitational solutions? Very little, surely.  There is a wide separation of scales between, say, the Compton wavelength of the electron and the curvature length scale outside of a typical black hole.  Particle physics is largely irrelevant to the study of gravitational fields produced by macroscopic objects.  However, while this hierarchy makes the gravitational signatures of heavy fields rather small, it also makes their study amenable to a controlled, local effective field theory (EFT) treatment.  In the present paper, we carry out this analysis and explore how high energy (UV) physics imprints itself on familiar, low energy (IR) gravitational solutions.

The role of heavy fields is explored in the context of static black holes with flat, de Sitter or anti-de Sitter asymptotics.  These Schwarzschild-$(A)dS$ spacetimes are the most fundamental, non-trivial solutions of pure Einstein Gravity coupled to a cosmological constant\footnote{Actions are denoted by $I$, while $S$ is reserved for entropy.  Anti-de Sitter, de Sitter, Schwarzschild-$AdS$ and Schwarzschild-$dS$ are abbreviated as $AdS$, $dS$, $SAdS$ and $SdS$, respectively.} (CC):
\begin{align}
I_{\rm GR}&=\int\rd^{4}x\sqrt{-g}\,\left (\frac{M_{p}^{2}}{2} R-\Lambda\right )\nn
\bar{g}_{\mu\nu}\rd x^{\mu}\rd x^{\nu}&\equiv-\left (1-\frac{r_{s}}{r}-\frac{\Lambda r^{2}}{3 M_{p}^{2}}\right )\rd t^{2}+\left (1-\frac{r_{s}}{r}-\frac{\Lambda r^{2}}{3 M_{p}^{2}}\right )^{-1}\rd r^{2}+r^{2}\rd \Omega^{2} \ .\label{GRActionAndSchwarzschildSolution}
\end{align}
Our goal is to determine how the introduction of heavy states alters the above family of solutions.

 We address this by adding massive fields to the GR action, $I_{\rm GR}\to I_{\rm GR}+I_{\rm M}$, and integrating them out.  The result is a specific higher derivative theory of the metric in which we can look for new black hole (BH) solutions.  As long as one studies black holes which are much larger than the Compton wavelength of the matter (e.g. $r_{s}\gg m^{-1}$ for pure Schwarzschild), the new solution will simply be \eqref{GRActionAndSchwarzschildSolution} plus tiny corrections.  
 
 We examine three simple cases in which we integrate out a massive, non-self interacting, minimally coupled scalar, Dirac fermion and vector field.  Each of these calculations is one-loop and hence we will be studying the \textit{quantum} corrections to the metric due to heavy fields. The calculations must be carried out to \textit{sixth} order in derivatives, as it's these operators, such as $R^{\rho\sigma\mu\nu}R_{\mu\nu\kappa\lambda}R^{\kappa\lambda}{}_{\rho\sigma}$ which provide the leading corrections to the solution.  Matter's contributions to the $M_{p}$ and $\Lambda$ beta functions are also calculated, as well as those of the four-derivative $\sim R^{2}$ operator coefficients.  However, the effect of these latter terms on the solution \eqref{GRActionAndSchwarzschildSolution} is subleading, a special property of $d=4$.
 
 After perturbatively solving the equations of motion, the changes to horizon radii, Hawking temperatures and entropies are found (in various limits) as well as the change in the energy of large $AdS$ black holes, as calculated from the first law.   In particular, we use the replica trick to efficiently calculate the entropy for static, spherically symmetric solutions of the most general four-dimensional gravitation action containing up to six derivatives of the metric.  An explicit demonstration of the method is given in Appendix \ref{Appendix:dSReplicaTrick} where we calculate the entropy of $dS_{4}$.

  There are several basic questions we can address, such as whether there's any universality in the way the matter affects solutions.  Do heavy fields seem to always push horizon radii inwards or outwards?  Or, do they always increase the Hawking temperature and entropy of black holes and cosmological horizons?
  
  Perhaps contrary to expectations, we find very few general patterns across particle species.  At fixed $r_{s}$, scalars and fermions decrease the horizon radii of all black holes, while vectors lead to an increase.   Scalars increase the temperature and entropy of all black hole horizons, while fermions decrease these quantities and the effect of vectors depends on the particular solution of interest (again, at fixed $r_{s}$).  For asymptotically de Sitter spacetimes, scalar fields shrink the cosmological horizon and increase its Hawking temperature and entropy, while fermions and vectors have exactly the opposite effect.    With entropy bounds in mind, we also compare the change in black hole entropy for solutions with fixed horizon area.  It's found that scalars and fermions increase the entropy one can store in a given region, as expected, but vectors strangely lead to a reduction.
  
  However, we also find that many of the corrections to black holes are so small as to be untrustworthy.  All effective field theories come equipped with a cutoff which sets the resolving power of the theory and determines its regime of validity.  In the present paper this is set by the distance scale $\sim m^{-1}$, the Compton wavelength of the heavy particle, and we find that various properties of the black hole are corrected at parametrically \textit{smaller} scales than this.  For instance, for every horizon we consider, the higher derivative operators shift the wavelength of the corresponding Hawking radiation by an amount $\delta \lambda \ll m^{-1}$.  Similarly, the proper areas of all horizons are changed by amounts $\delta A\ll m^{-2}$.  Therefore, neither correction is credible, from an EFT perspective.   Many fundamental properties of gravitational solutions therefore appear remarkably robust against higher derivative gravitational corrections.

The paper is structured as follows.  In Sec.$\!$ \ref{Sec:GenericHigherDerivativeCalculations} we consider a generic higher derivative, gravitational action containing all operators with up to six derivatives of the metric with arbitrary coefficients and use it to calculate the leading corrections to the solution \eqref{GRActionAndSchwarzschildSolution}.  In Sec.$\!$ \ref{Sec:EffectiveActions} we calculate the actual values of these coefficients which arise when scalars, Dirac fermions and vectors are integrated out.  In Sec.$\!$ \ref{Sec:AllAsymptoticPossibilities} we study the general solution of Sec.$\!$ \ref{Sec:GenericHigherDerivativeCalculations} evaluated on the specific coefficients found in Sec.$\!$ \ref{Sec:EffectiveActions} and in Sec.$\!$ \ref{Sec:Interpretation} we provide some interpretation and discussion of the results found in the preceding section.  In Sec.$\!$ \ref{Sec:ValidityEFT} we evaluate the size of the corrections from an EFT point of view, demonstrating that many are too small to trust. In Sec.$\!$ \ref{Sec:Conclusion} we conclude.  Finally, we should note that similar ideas have been previously explored in, for instance, \cite{Ruffini:2013hia,Lu:1993sq,Matyjasek:2006fq,Taylor:1999ic,Kats:2006xp,Smolic:2013gz} and especially \cite{Matyjasek:2006nu} which considers the effects of massive scalars, fermions and vectors on the Reissner-Nordstrom solution.  This paper can be viewed as an extension of \cite{Matyjasek:2006nu} to spacetimes with non-flat asymptotics.

\textbf{Conventions:} 	Our metric and curvature conventions are those of Carroll \cite{Carroll:2004st} (equivalently, Misner, Thorne and Wheeler \cite{Misner:1974qy}): we work in mostly \textit{plus} signature, $\eta_{\mu\nu}=(-,+,+,+)$ and use the curvature conventions
\begin{align}
\Gamma_{\mu\nu}^{\lambda}&= \frac{1}{2}g^{\lambda\sigma}\left [\partial_{\mu }g_{\nu\sigma}+\partial_{\nu}g_{\sigma\mu}-\partial_{\sigma}g_{\mu\nu}\right ]\ , \quad
\left [\nabla_{\mu},\nabla_{\nu}\right ]V^{\rho}=R^{\rho}{}_{\sigma\mu\nu}V^{\sigma}\nn
R^{\rho}{}_{\sigma\mu\nu}&=\partial_{\mu } \Gamma^{\rho}_{\nu\sigma}-\partial_{\nu} \Gamma^{\rho}_{\mu\sigma}+ \Gamma^{\rho}_{\mu\lambda}\Gamma^{\lambda}_{\nu\sigma}-\Gamma^{\rho}_{\nu\lambda}\Gamma^{\lambda}_{\mu\sigma}\ , \quad
R_{\sigma\nu}=R^{\rho}{}_{\sigma\rho\nu}\ .
\end{align} The number of spacetime dimensions is $d$, though we nearly always work in $d=4$.  The Planck mass conventions are $M_{p}^{2}\equiv 1/l_{p}^{2}\equiv \left (8\pi G_{N}\right )^{-1}$ so that the Schwarzschild radius for a black hole of mass $M$ is $r_{s}\equiv\frac{M}{4\pi M_{p}^{2}}$. The $(A)dS$ radius is $L_{(A)dS}\equiv\sqrt{3M_{p}^{2}/|\Lambda|}$ and when considering cases with a definite sign on $\Lambda$ we will replace factors of $\Lambda$ by $L_{(A)dS}$.

\section{Generic Higher Derivative Calculations\label{Sec:GenericHigherDerivativeCalculations}}

We begin by deriving general formulas for how higher derivative operators perturb the solution \eqref{GRActionAndSchwarzschildSolution}. We find the leading corrections from the most general gravitational action containing up to six derivatives with arbitrary coefficients in front of each independent operator. In later sections the coefficients are replaced by the specific values which arise when particular forms of matter are added to the action.  It will turn out that the $\sim R^{3}$ operators  provide the leading corrections to the solution; neither the $\sim R^{2}$ or $\sim R\nabla\nabla R$ operators contribute, at lowest order. 

\subsection{The Generic Action}

Generically, when a heavy field is integrated out, it generates a tower of operators which have an expansion in $R_{\mu\nu\rho\sigma}/m^{2}$ and $\nabla^{2}/m^{2}$ so that the (renormalized) gravitational action reads
\begin{align}
I_{\rm eff}&\sim \int\rd^{4}x\, \sqrt{-g}\Big[\frac{M_{p}^{2}}{2}R-\Lambda+ a R^{2}+\frac{b}{m^{2}} R\nabla\nabla R+\frac{c}{m^{2}}R^{3}+\ldots\Big]\ ,\label{SchematicEffectiveAction}
\end{align}
schematically, where all indices have been suppressed, $a,b,c$ are $\mathcal{O}(1)$ coefficients and the ellipses contains operators with eight or more derivatives.  Minimally coupled matter generates divergent contributions to each classically relevant or marginal operator in \eqref{SchematicEffectiveAction}, which need to be regularized and renormalized in the usual manner.

  Some comments:
\begin{itemize}
\item Above, we've implicitly tuned the bare values of $M_{p}$ and $\Lambda$ in the theory including heavy matter so that their renormalized values appearing in \eqref{SchematicEffectiveAction} are the same as their counterparts in the original GR action \eqref{GRActionAndSchwarzschildSolution}.  This makes comparisons between the original and corrected solutions clean as it ensures that solutions of \eqref{SchematicEffectiveAction} smoothly match onto those of \eqref{GRActionAndSchwarzschildSolution} in the limit that the matter becomes infinitely heavy and decouples, $m\to \infty$.  This choice is discussed in some more detail in Sec.$\!$ \ref{Sec:EffectiveActionsIntegratingOut} and revisited later in Sec.$\!$ \ref{Sec:EntropyDiscussion}.
\item 	The natural size of the CC is $|\Lambda|\gtrsim\mathcal{O}(m^{4})$ (at least), but we treat $\Lambda$ as an arbitrary parameter.
\item Integrating out heavy fields generates divergent $\sim R^{2}$ terms, hence we need to include such operators in the bare action with bare coefficients.  The renormalized values of these coefficients are therefore incalculable and we just assume they are $\mathcal{O}(1)$, their natural size.  Fortunately, such terms do not change the solution at leading order; see Sec. \ref{Sec:R2Terms}.
\item The remaining coefficients in \eqref{SchematicEffectiveAction} are finite and unambiguous.
\item The action is an infinite expansion in powers of $\sim R/m^{2}$ and $\nabla^{2}/m^{2}$, hence our perturbative solution will only be trustworthy if we study regimes where these terms are small. In the cases we consider, this translates into the conditions
\begin{align}
r_{s}\gg m^{-1}\ , \quad |\Lambda|^{-1}M_{p}^{2} \ll m^{-2}\ .\label{EFTValidityConditions}
\end{align}
Physically, both the black hole and the $(A)dS$ length scale $L_{(A)dS}\sim \left (|\Lambda|^{-1}M_{p}^{2}\right )^{1	/2}$ need to be much larger than the Compton wavelength of the heavy particle. Otherwise, we're not working at length scales where heavy particle physics is negligible (for instance, we might expect pair production to become rapid when the conditions in \eqref{EFTValidityConditions} are violated).  Of course, we also assume $m^{-1}\gg l_{p}$ throughout, implying the hierarchy $L_{(A)dS},r_{s}\gg m^{-1}\gg l_{p}$. 
\end{itemize}

We now set the conventions for the precise higher derivative theory under study.  The action is written as $I=I_{\rm GR}+I_{\rm EFT}^{\partial^{4}}+I_{\rm EFT}^{\partial^{6}}+\ldots$ where
\begin{align}
I_{\rm GR}&\equiv \int\rd^{4}x\, \sqrt{-g}\left(\frac{M_{p}^{2}}{2}R-\Lambda\right )\nn
I_{\rm EFT}^{\partial^{4}}&=\int\rd^{4}x\,\sqrt{-g}\left (a_{1}R^{2}+a_{2}R_{\mu\nu}^{2}+a_{\rm GB}\mathcal{L}_{\rm GB}\right )\nn 
I_{\rm EFT}^{\partial^{6}}&\equiv \int\rd^{4}x\sqrt{-g}\,m^{-2} \Big[b_{1}R\square R+b_{2}R^{\mu\nu}\square R_{\mu\nu}+c_{1}R^{\rho\sigma\mu\nu}R_{\mu\nu\kappa\lambda}R^{\kappa\lambda}{}_{\rho\sigma}+c_{2}RR^{\mu\nu\rho\sigma}R_{\mu\nu\rho\sigma} \nn
&\quad +c_{3}R^{3}+c_{4}R^{\mu }{}_{\nu}R^{\nu}{}_{\rho}R^{\rho}{}_{\mu}+c_{5}RR^{\mu\nu}R_{\mu\nu}  +c_{6}R^{\mu\rho}R^{\nu\sigma}R_{\mu\nu\rho\sigma}\Big]\label{PreciseAction}\ ,
\end{align}
where $\mathcal{L}_{\rm GB}$ is the Gauss-Bonnet term
\begin{align}
\mathcal{L}_{\rm GB}&\equiv R^{2}-4R_{\mu\nu}R^{\mu\nu}+R_{\mu\nu\rho\sigma}R^{\mu\nu\rho\sigma}\label{GaussBonnet}\ .
 \end{align}

 The action \eqref{PreciseAction} forms a basis for all local operator up to sixth order in derivatives in $d=4$, after taking into account integrations by parts, Bianchi identities and the dimension dependent identities \cite{Edgar:2001vv}
 \begin{align}
 0&= R^{\mu }{}_{[\mu }{}^{\nu}{}_{\nu}R^{\rho}{}_{\rho}{}^{\sigma}{}_{\sigma}R^{\kappa}{}_{\kappa}{}^{\lambda}{}_{\lambda]}  \ , \quad  {\rm and} \quad \ 0=R^{\mu }{}_{[\mu }{}^{\nu}{}_{\nu}R^{\rho}{}_{\rho}{}^{\sigma}{}_{\sigma}R^{\kappa}{}_{\kappa]}\label{DimensionDependentIdentitiesExplicit}\ .
 \end{align}  See \cite{Fulling:1992vm} for  a study of bases of higher derivative operators in various dimensions (not accounting for integrations by parts).

\subsection{Metric Solution}

Here we use the action \eqref{PreciseAction} to find perturbative corrections to the solution \eqref{GRActionAndSchwarzschildSolution}.

\subsubsection{Perturbative Expansion}

The gravitational equations of motion (EOM) are solved perturbatively.  
  Splitting the effective action into its GR and EFT components \eqref{PreciseAction}, $I_{\rm GR}$ and $I_{\rm EFT}\equiv I_{\rm EFT}^{\partial^{4}}+I_{\rm EFT}^{\partial^{6}}+\ldots$ we treat the latter terms as $\mathcal{O}(\hbar)$ corrections to $I_{\rm GR}$ and find solutions to linear order in $\hbar$.  In practice, this means we work to first order in the $\{a_{i},b_{i},c_{i}\}$ EFT coefficients \eqref{PreciseAction}, as we mostly set $\hbar=1$.
  
  The equations of motion for the metric $g_{\mu\nu}$ can be written as
\begin{align}
\mathcal{E}_{\mu\nu}[g]&= \frac{1}{M_{p}^{2}}T_{\mu\nu}^{\rm EFT} [g]\ , \label{EFTEOM}
\end{align}
where $\mathcal{E}_{\mu\nu}[g]$ is the familiar GR equation of motion and $T_{\mu\nu}^{\rm EFT}$ is the effective stress tensor from the EFT operators:
\begin{align}
\mathcal{E}_{\mu\nu}[g]&\equiv G_{\mu\nu}+\frac{\Lambda}{M_{p}^{2}}g_{\mu\nu}\nn
T_{\mu\nu}^{\rm EFT}[g]&\equiv	-\frac{2}{\sqrt{-g}}\frac{\delta I_{\rm EFT}}{\delta g^{\mu\nu}}\label{EandTmunuEFTDefinitions}\ .
\end{align}

When used in their regime of validity, the EFT operators only generate slight deviations from the Schwarzschild-$(A)dS$ solution, so \eqref{EFTEOM} can be solved perturbatively as an expansion about \eqref{GRActionAndSchwarzschildSolution}
\begin{align}
g_{\mu\nu}&=\bar{g}_{\mu\nu}+\delta g_{\mu\nu}
\end{align}
where $\delta g_{\mu\nu}$ is linear in the EFT coefficients.  The first order correction is then found by evaluating $T^{\rm EFT}_{\mu\nu}$ on the background solution, expanding $\mathcal{E}_{\mu\nu}[\bar{g}+\delta g]$ to linear order in the fluctuation and solving for $\delta g_{\mu\nu}$.  Note what while we are working to linear order in EFT coefficients, we are working non-linearly in all other factors.  For instance, when we find corrections to the flat Schwarzschild solution, we will work all orders in $r_{s}/r$ as is necessary to reliably explore the near horizon regime.

\subsubsection{ $R^{2}$ Terms\label{Sec:R2Terms}}

The $\sim R^{2}$ operators in $I_{\rm EFT}^{\partial^{4}}$ come with coefficients which are not unambiguously determined by the properties of the heavy matter in the theory.  Loops of both heavy matter fields and gravitons themselves \cite{'tHooft:1974bx} generate divergent contributions to these coefficients which only tell us that the natural size of their coefficient is $\mathcal{O}(1)$.  These are also the leading terms in the derivative expansion and hence they could be expected to generate the leading corrections the metric.   It is then a fortunate property of $d=4$ that the contribution of $\sim R^{2}$ operators to $T_{\mu\nu}^{\rm EFT}$ is vanishing, at leading order.

First, the Gauss-Bonnet combination would contribute non-trivially in $d>4$, but becomes a total derivative\footnote{Though $\mathcal{L}_{\rm GB}$ doesn't enter $T^{\rm EFT}_{\mu\nu}$, and hence doesn't alter the solution, it \textit{does} affect black hole entropy \cite{Liko:2007vi,Chatterjee:2013daa,Sarkar:2010xp}, so it's kept in the action. } in four dimensions.  The operators $R^{2}$ and $R_{\mu\nu}^{2}$ form a basis the remaining possible four-derivative terms, but their contribution to $T_{\mu\nu}^{\rm EFT}$ is also vanishing in $d=4$.  Working in $d$-dimensions, the contribution of $I_{\rm EFT}^{\partial^{4}}$ to the effective stress tensor is:
   \begin{align}
     T_{\mu\nu}^{\rm EFT}&\supset \frac{4(d-4)(d\,a_{1}+a_{2})\Lambda^{2}}{(d-2)^{2}M_{p}^{4}}\bar{g}_{\mu\nu}\ ,\label{VanishingContributionOfAOperatorsin4D}
     \end{align}
     after simplifying using identities and background equations of motion.
     It is therefore a special property of  $d=4$ that $I_{\rm EFT}^{\partial^{4}}$ does not contribute to $T_{\mu\nu}^{\rm EFT}$ for arbitrary $\Lambda$, at first order (they never contribute in any $d$ for the special case $\Lambda=0$, due to Ricci flatness of the background).   This can be understood as a consequence of the scale invariance of $\sim R^{2}$ terms in four dimensions.

\subsubsection{$R^{3}$ and $R\nabla\nabla R$ Terms\label{Sec:R3AndRdeldelRterms}}

Next are terms of the form $\sim R\nabla\nabla R$ and $\sim R^{3}$ in $I_{\rm EFT}^{\partial^{6}}$, each of which comes with an unambiguous coefficient. Evaluating these terms on the background \eqref{GRActionAndSchwarzschildSolution} to find $T_{\mu\nu}^{\rm EFT}$ is straightforward, if tedious. The relevant equations of motion can be easily derived and simplified using the excellent \textit{xAct} \cite{xAct} package.

Performing the calculation, the non-trivial stress tensor components are found to be:
\begin{align}
T^{\rm EFT}_{tt}&=\frac{2\Delta(r)}{9m^{2}r^{9}}\Big[\left (-2646 c_1 -3564 c_2\right )r_{s}^{3}+ (2430 c_1 +3240 c_2)r r_s ^2+(-594 c_1 -756 c_2) \frac{r^3 r_s ^2   \Lambda}{M_p^2}\nn
&\quad+(4 c_1 +24 c_2+144 c_3+9 c_4+36
   c_5+9 c_6) \frac{r^9  \Lambda ^3}{M_p^6}\Big ]\nn
T^{\rm EFT}_{rr}&=-\frac{2}{9m^{2}r^{9}\Delta(r)}\Big[ (270 c_1 +972 c_2)r_s ^3+ (-486 c_1 -1296 c_2)r r_s ^2+(378 c_1 +756 c_2) \frac{r^3 r_s ^2  \Lambda}{M_p^2} \nn
&\quad+ (4 c_1 +24 c_2+144 c_3+9 c_4+36
   c_5+9 c_6)\frac{r^9 \Lambda ^3}{M_p^6} \Big ]\nn
T^{\rm EFT}_{\theta\theta}&=-\frac{2}{9m^{2}r^{7}}\Big[ (-1674 c_1 -4536 c_2)r_s ^3+(1458 c_1 +3888 c_2)r r_s ^2 + (-270 c_1 -756 c_2) \frac{r^3 r_s ^2 \Lambda}{M_p^2} \nn
&\quad+(4 c_1 +24 c_2+144 c_3+9 c_4+36
   c_5+9 c_6) \frac{r^9   \Lambda ^3}{M_p^6}\Big ]\nn
T^{\rm EFT}_{\phi\phi}&=T^{\rm EFT}_{	\theta\theta}\sin^{2}\theta \ , \quad \Delta(r)\equiv 1-\frac{r_{s}}{r}-\frac{\Lambda r^{2}}{3M_p^2}\label{ExplicitStressTensors}\ .
\end{align}
The above terms generate the leading corrections to the metric.  Further corrections arising from $\sim R^{4}$ operators, for instance, are suppressed relative to \eqref{ExplicitStressTensors} by factors of $(m r_{s})^{-2}$ and $(mL_{(A)dS})^{-2}$. 

Note that \eqref{ExplicitStressTensors} is independent of the $b_{i}$'s. 
The $\sim R\nabla\nabla R$ operators don't contribute to $T_{\mu\nu}^{\rm EFT}$, but each of the $c_{i}$ terms $\sim R^{3}$ do\footnote{The $b_{i}$'s never contribute in any dimension and only $c_{1},c_{2}$ contribute non-trivially in the Ricci flat $\Lambda\to 0$ limit.}. This is a leading order property which follows from the fact that for the background solution \eqref{GRActionAndSchwarzschildSolution} the Ricci tensor is directly proportional to the metric and hence covariantly constant.  The variations of both $R\square R$ and $R^{\mu\nu}\square R_{\mu\nu}$ only generate terms proportional to a factor of $\nabla_{\mu}R_{\nu\sigma}$ which vanishes, a nice aspect of working in the basis \eqref{PreciseAction}.  As a result, nearly all subsequent formulas in this paper are independent of the $b_{i}$'s.

\subsubsection{The Solution}

Given the above, solving for the perturbed metric is straightforward.  Writing the solution as 
\begin{align}
g_{\mu\nu}\rd x^{\mu}\rd x^{\nu}&\equiv-\left (1-\frac{r_{s}}{r}-\frac{\Lambda r^{2}}{3 M_{p}^{2}}+\delta g_{1}\right )\rd t^{2}+\left (1-\frac{r_{s}}{r}-\frac{\Lambda r^{2}}{3 M_{p}^{2}}+\delta g_{2}\right )^{-1}\rd r^{2}+r^{2}\rd \Omega^{2} \ ,\label{PerturbedMetricExplicit1}
\end{align}
understood to only be valid at first order in the $\delta g_{i}$ perturbations, we find:
\begin{align}
\delta g_{1}&= -\frac{\mathcal{C}_{1}r_{s}}{r}+\mathcal{C}_{2}\left (1-\frac{r_{s}}{r}-\frac{\Lambda r^{2}}{3M_{p}^{2}}\right )+(10 c_1 +36 c_2)\frac{r_s ^3  }{m^2 M_p^2r^7}-24 c_2 \frac{ r_s ^2}{m^2M_p^2 r^6}-8 c_1 \frac{ r_s ^2 \Lambda  }{m^2 M_p^4 r^4}\nn
&\quad+\left(-\frac{8 c_1 }{27}-\frac{16
   c_2}{9}-\frac{32 c_3}{3}-\frac{2 c_4}{3}-\frac{8
   c_5}{3}-\frac{2 c_6}{3}\right)\frac{r^2 \Lambda ^3}{m^2 M_p^8}\nn
   \delta g_{2}&= -\frac{\mathcal{C}_{1}r_{s}}{r}+(-98 c_1 -132 c_2)\frac{r_s ^3  }{m^2 M_p^2r^7}+(108 c_1 +144 c_2)\frac{ r_s ^2 }{m^2M_p^2 r^6}+(-44 c_1 -56 c_2)\frac{r_s ^2   \Lambda}{m^2M_p^4 r^4}\nn
   &\quad+\left(-\frac{8 c_1 }{27}-\frac{16
   c_2}{9}-\frac{32 c_3}{3}-\frac{2 c_4}{3}-\frac{8
   c_5}{3}-\frac{2 c_6}{3}\right)\frac{ r^2   \Lambda ^3}{m^2M_p^8}   \label{PerturbedMetricExplicit2} \ .
\end{align}
Higher derivative corrections spoil the equivalence between $-g_{tt}$ and $ g_{rr}^{-1}$ present in the background solution \eqref{GRActionAndSchwarzschildSolution}.

Above, $\mathcal{C}_{1}$ are $\mathcal{C}_{2}$ are undetermined integration constants which are $\mathcal{O}(\hbar)$, if $\hbar$'s are restored. The coefficient $\mathcal{C}_{1}$ clearly corresponds to the arbitrary nature of choosing the Schwarzschild radius of the black hole, so we mostly set $\mathcal{C}_{1} \to 0$ in this paper. For asymptotically flat black holes, this ensures that the mass of the black hole in the higher derivative theory is the same as that of the background GR solution\footnote{\label{Foot:MisEFootnoteFlatSpace}This essentially follows from dimensional analysis.  For GR, the ADM mass arises as $M_{\rm ADM}\sim \int\rd^{2}\sigma\, M_{p}^{2}(K-K_{\rm flat})$.  In the higher derivative theory, the appropriate generalized ADM mass expression should additionally contain terms $\Delta M_{\rm ADM}\sim\int\rd^{2}\sigma\,\left (a_{i}A_{i}+b_{i}B_{i}/m^{2}+c_{i}C_{i}/m^{2}\right ) $ with $a_{i},b_{i},c_{i}$ the dimensionless EFT coefficients in \eqref{PreciseAction} and $A_{i},B_{i},C_{i}$ functions of intrinsic and extrinsic curvatures.  Dimensional analysis (or explicit computation) demonstrates that $A_{i}, B_{i}, C_{i}$ must all fall off faster than $\sim 1/r^{2}$ and hence $\Delta M_{\rm ADM}$ simply vanishes.  Similarly, higher derivative corrections to $K$ (and $K_{\rm flat})$ also fall off too quickly to affect the original $M_{\rm ADM}$ integral.  The argument doesn't apply to asymptotically $AdS$ spaces where higher derivative operators \textit{do} alter the energy, see Sec.$\!$ \ref{Sec:FirstLaw}.  }.  In Sec.$\!$ \ref{Sec:EntropyDiscussion}, however, we briefly discuss entropy bounds and hence are interested in comparing black holes of fixed horizon area, in which case we effectively take $\mathcal{C}_{1}$ to be non-zero.    Meanwhile, $\mathcal{C}_{2}$ arises from the freedom to rescale time and is removed by sending $t\to (1-\mathcal{C}_{2}/2)t$, treating $\mathcal{C}_{2}$ perturbatively.  See Sec.$\!$ 3.3.1 of \cite{Smolic:2013gz} for related discussions on choosing integration constants in higher derivative gravity

Higher derivative operators also alter the coefficient of the $r^{2}$ terms in $g_{tt}$ and $g_{rr}^{-1}$, producing an effective shift to the cosmological constant $\Lambda\to \Lambda_{\rm eff}$ where
\begin{align}
\frac{\Lambda_{\rm eff}}{\Lambda}&=1+ \left(\frac{8 c_1 }{9}+\frac{16 c_2}{3}+32
   c_3+2 c_4+8 c_5+2 c_6\right)\frac{ \Lambda ^2 }{m^2M_p^6}\ .\label{LambdaEff}
\end{align}
The fractional correction has the same sign for either $dS$ or $AdS$ and is small when the EFT is used in its regime of validity \eqref{EFTValidityConditions}.

\subsection{Horizons, Hawking Temperatures and Entropies}

Given the solution \eqref{PerturbedMetricExplicit2}, we can find how horizons, Hawking temperatures and entropies are changed by heavy matter.

\subsubsection{Horizons}

We now study how horizons are shifted in various limiting cases.  

If a horizon occurs at $r=\bar{r}_{h}$ in the background metric \eqref{GRActionAndSchwarzschildSolution}, the higher derivative corrections shift it to $r_{h}=\bar{r}_{h}+\delta r$.  Substituting into \eqref{PerturbedMetricExplicit1} and solving perturbatively, we find
\begin{align}
\delta r&=-\frac{M_p^2}{9 \Lambda   \bar{r}_{h}^8 \left(1-\frac{3 r_s M_p^2 }{2 \Lambda 
   \bar{r}_{h}^3}\right)}\Big[(-135 c_1 -486 c_2)\frac{r_s ^3  }{m^2 M_p^2}+324 c_2 \frac{ r_s ^2 \bar{r}_{h} }{m^2M_p^2}+108 c_1\frac{   r_s ^2 \bar{r}_{h}^3  \Lambda }{m^2M_p^4}\nn
   &\quad+(4 c_1 +24 c_2+144 c_3+9
   c_4+36 c_5+9 c_6)\frac{ \bar{r}_{h}^9   \Lambda ^3}{m^2 M_p^8}\Big]\ .\label{deltarSolution}
\end{align}
We now evaluate \eqref{deltarSolution} in various scenarios.

The following cases are simple limits of \eqref{deltarSolution}:
\begin{itemize}
\item BH horizon with $r_{s}^{2}l_{p}^{2}|\Lambda|\ll 1$, i.e.$\!$ small black holes $r_{s}\ll L_{(A)dS}$.  These exist for both $AdS$ and $dS$ asymptotics and the preceding condition implies that the cosmological and black hole horizons are far separated in the latter case.  We find:
\begin{align}
\bar{r}_{h}&=r_{s}+\frac{1}{3}r_{s}^{3}l_{p}^{2}\Lambda+\ldots\nn
\frac{\delta r}{\bar{r}_{h}}&=(-10 c_1 -12 c_2)\left (\frac{l_p ^2 }{m^2 r_s ^4}\right )+\ldots\ ,\label{BHHorizonShiftSmallRs}
\end{align}
where the ellipses contains terms suppressed by powers of $r_{s}/L_{(A)dS}\ll 1$.
\item BH horizon for $r_{s}^{2}l_{p}^{2}|\Lambda|\gg 1$, i.e.$\!$ large black holes $r_{s}	\gg L_{AdS}$.  As indicated, these are only sensible for $AdS$ asymptotics. We find:\begin{align}
\bar{r}_{h}
&=\left (r_{s}L_{AdS}^{2}\right )^{1/3}-\frac{1}{3}\left (r_{s}L_{AdS}^{2}\right )^{1/3}\left (\frac{L_{AdS}}{r_{s}}\right )^{2/3}\ldots\nn
\frac{\delta r}{\bar{r}_{h}}&=(-14 c_1 -28 c_2-96 c_3-6
   c_4-24 c_5-6 c_6)\left (\frac{l_{p}^{2}}{m^{2}L_{AdS}^{4}}\right )+\ldots\ ,\label{BHHorizonShiftLargeRs}
\end{align}
where the ellipses contains terms suppressed by powers of $L_{AdS}/r_{s}\ll 1$.
\item The cosmological horizon for $r_{s}^{2}l_{p}^{2}\Lambda\ll 1$, i.e.$\!$ small Schwarzschild-de Sitter black holes, $r_{s}\ll L_{dS}$.  We find:
\begin{align}
\bar{r}_{h}&=L_{dS}-\frac{r_{s}}{2}+\ldots\nn
\frac{\delta r}{\bar{r}_{h}}&=(-4 c_1 -24 c_2-144 c_3-9 c_4-36 c_5-9
   c_6)\left (\frac{l_p ^2 }{m^2L_{dS}^4 }\right )+\ldots\label{CosHorizonShiftSmallRs}
\end{align}
where the ellipses contains terms suppressed by powers of $r_{s}/L_{dS}\ll 1$.
\end{itemize}
Above, we've traded mass scales for length scales when convenient and given fractional corrections to emphasize the perturbative nature of the calculation: $\delta r/\bar{r}_{h}\ll 1$ when the EFT description is valid \eqref{EFTValidityConditions}.

A final interesting limit, which requires more detailed comment, are black holes in $dS$ whose BH and cosmological horizons are approaching one another: $r_{s}\approx L_{dS}$.  Naively, this case is problematic as \eqref{deltarSolution} diverges in this regime since $\left(1-\frac{3 r_s }{2 \Lambda 
   l_p ^2 \bar{r}_{h}^3}\right)$ tends towards zero, vanishing precisely when the two horizons merge, yielding the Nariai metric:
\begin{align}
\bar{g}_{\mu\nu}^{\rm Nariai}&=-\Delta_{N}(r)\rd t^{2}+\Delta_{N}(r)^{-1}\rd r^{2}+r^{2}\rd \Omega^{2} \ , \quad
\Delta_{N}(r)\equiv \frac{r_{s}}{r}\left (1+\frac{r}{3r_{s}}\right )\left (1-\frac{2r}{3r_{s}}\right )^{2}\  .
\end{align}
This solution corresponds to \eqref{GRActionAndSchwarzschildSolution} with $ \Lambda=\frac{4}{9l_{p}^{2}r_{s}^{2}}$.

Because of the double pole structure of $\Delta_{N}(r)$ at the single horizon $r=\frac{3}{2}r_{s}$, the shift $\delta r$ to the horizon induced by higher order terms is proportional to the \textit{square root} of the $c_{i}$ coefficients, rather than linear in $c_{i}$'s as was implicitly assumed in deriving \eqref{deltarSolution}.   Accounting for this rectifies the divergence in the naive formula \eqref{deltarSolution}. Similar behavior occurs when finding the corrections to extremal Reissner-Nordstom black holes due to QED effects \cite{ToAppearWithKurt}.

 It is straightforward to perturbatively solve for the shifted horizon for near-Nariai solutions.  Parameterizing the cosmological constant by $\Lambda=\frac{4}{9l_{p}^{2}r_{s}^{2}}(1+\lambda)$, $|\lambda|\ll 1$, and writing the corrected horizon as $r_{h}=\bar{r}_{h}+\delta r$ with $\bar{r}_{h}=\frac{3}{2}r_{s}$, we find
\begin{align}
\delta r=\pm \frac{3}{2}r_{s}\sqrt{ -\frac{1}{729}\left (128 c_1 +256 c_2+1536 c_3+96 c_4+384 c_5+96 c_6\right )\left (\frac{l_{p}}{mr_{s}^{2}}\right )^{2}-\frac{\lambda}{3}}\ .\label{NariaiHorizonShift}
\end{align}
If the net contribution of the $c_{i}$ terms in the argument of \eqref{NariaiHorizonShift} is positive, then heavy fields have caused the BH and cosmological horizons to separate from each other (for fixed $\lambda$).  If the net contribution is negative, then heavy fields cause the horizons to push towards each other and one needs to take $\lambda<0$ in order to prevent the horizons from passing through each other and generating a naked singularity.  We will not consider these near-Nariai black holes again in this paper, focusing instead on the other three limits presented in this section. See \cite{Ginsparg:1982rs,Bousso:1997wi,Nojiri:1998ph,Bytsenko:1998md,Nojiri:2013su} for interesting discussions of Nariai physics.

\subsubsection{Hawking Radiation\label{Sec:HawkingRadiation}}

The Hawking radiation associated to the various horizons is easily calculated via standard Euclidean gravity methods \cite{Gibbons:1976ue}. 

 The solution \eqref{PerturbedMetricExplicit1} is of the form
\begin{align}
\rd s^{2}&=-f(r)\rd t^{2}+\frac{1}{g(r)}\rd r^{2}+r^{2}\rd \Omega^{2}\ \label{GenericSphericalMetric}
\end{align}
and at each horizon, $r=r_{h}$, both $f(r_{h})$ and $g(r_{h})$ vanish\footnote{Though $f(r)\neq g(r)$ for generic $r$, their zeros coincide.  This must occur for the spacetime to be regular.}.  We Euclideanize $t\to -i\tau$ and probe the near horizon limit by letting $r=r_{h}+\frac{g'(r_{h})}{4}\rho^{2}$, yielding
\begin{align}
\rd s^{2}&=\frac{f'(r_{h})g'(r_{h})}{4}\rho^{2}\rd \tau^{2}+\rd \rho	^{2}+r_{h}^{2}\rd \Omega^{2}+\ldots \label{NearHorizonLimitGeneric}
\end{align}
A conical singularity at $\rho=0$ is avoided only if we identify
\begin{align}
\tau\sim \tau+\frac{4\pi}{\sqrt{f'(r_{h})g'(r_{h})}}
\end{align}
and this periodicity determines the Hawking temperature of the horizon
\begin{align}
T_{H}&\equiv\beta^{-1}=\frac{\sqrt{f'(r_{h})g'(r_{h})}}{4\pi}\ .\label{GenericHawkingTemperature}
\end{align}

Now we evaluate the Hawking temperature \eqref{GenericHawkingTemperature} for the three cases emphasized in the previous section.  We denote the background value of the temperature by $\bar{T}_{H}$ and write the corrected value as $T_{H}=\bar{T}_{H}+\delta T_{H}$. We have:
\begin{itemize}
\item For small black holes $r_{s}\ll L_{(A)dS}$, the BH horizon generates a temperature
\begin{align}
\bar{T}_{H}&=\frac{1}{4 \pi  r_s }-\frac{\Lambda  l_p ^2 r_s }{3 \pi }+\ldots\nn
\frac{\delta T_{H}}{\bar{T}_{H}}&=4c_1\left ( \frac{ l_p ^2 }{  m^2 r_s ^4}\right )+\ldots\ ,\label{THShiftSmallRs}
\end{align}
where the ellipses contains terms suppressed by powers of $r_{s}/L_{(A)dS}\ll 1$.  
\item For large $SAdS$ black holes $r_{s}	\gg L_{AdS}$, the BH horizon generates a temperature
\begin{align}
\bar{T}_{H}&=\frac{3 }{4 \pi  L_{AdS}}\left (\frac{r_{s}}{L_{AdS}}\right )^{1/3}+\frac{1}{12\pi r_{s}}\ldots\nn
\frac{\delta T_{H}}{\bar{T}_{H}}&=(4 c_1+32 c_2+192 c_3 +12 c_4+48 c_5+12
   c_6)\left (\frac{l_p ^2 }{ m^2L_{AdS}^4}\right )+\ldots\ ,\label{THShiftLargeRs}
\end{align}
where the ellipses contains terms suppressed by powers of $L_{AdS}/r_{s}\ll 1$.
\item For small $SdS$ black holes, $r_{s}\ll L_{dS}$, the cosmological horizon generates a temperature
\begin{align}
\bar{T}_{H}&=\frac{1}{2 \pi  L_{dS}}-\frac{r_s }{2 \pi  L_{dS}^2}+\ldots\nn
\frac{\delta T_{H}}{\bar{T}_{H}}&=(4 c_1+24 c_2+144 c_3 +9 c_4+36 c_5+9
   c_6)\left (\frac{l_p ^2 }{m^2L_{dS}^4 }\right )+\ldots\label{THShiftCos}
\end{align}
where the ellipses contains terms suppressed by powers of $r_{s}/L_{dS}\ll 1$.  
\end{itemize}
Above, we've traded mass scales for length scales when convenient and given fractional corrections to emphasize the perturbative nature of the calculation: $\delta T_{H}/\bar{T}_{H}\ll 1$ when the EFT description is valid \eqref{EFTValidityConditions}: $L_{(A)dS}, r_{s}\gg m^{-1}\gg l_{p}$.  Note that in the large $AdS$ BH scenario, we should impose the further restriction $r_{s}\ll L_{ AdS}(mL_{AdS})^{3}$, so that the Hawking temperature is much smaller than $m$.

Before moving on to entropy calculations, we should mention that Euclidean calculations in Schwarzschild-$dS$ are somewhat dubious.  This is because preventing the existence of conical singularities at the black hole and cosmological horizons requires two \textit{different} periodicities of $\tau$, generically.  This reflects the fact that the black hole and cosmological horizon will not typically be in equilibrium with each other: the Hawking temperature associated to the black hole horizons is usually much larger than that of the cosmological horizon, meaning that the black hole will not typically be in thermal equilibrium with the cosmological heat bath.  Only in the Nariai limit do the two temperatures coincide \cite{Bousso:1997wi}.  Nevertheless, we will continue to work with Euclideanized $SdS$ spacetimes as they (at least) make sense in the $r_{s}\to 0$ or $\Lambda\to 0$ limits.

\subsubsection{Review: Entropy from the Replica Trick}

Finally, we can calculate the entropy of the various horizons. The calculation can be done in many ways, for instance via Wald's formula and refinements thereof \cite{Wald:1993nt,Iyer:1994ys,Jacobson:1993vj}.  Here, we instead employ the replica trick as presented by Lewkowycz and Maldacena  in Sec. 3 of \cite{Lewkowycz:2013nqa} (see references therein for earlier related work).

In  this section, we review the replica trick. This provides an extremely quick and efficient method for calculating the horizon entropy in generic theories of gravity.  In particular, it's easy to implement the method using computers.   In the following section, we present the results of the entropy calculation for generic, static $d=4$ horizons for the higher derivative theory \eqref{PreciseAction} and evaluate the results in the limiting cases we've been focusing on.

  Given a normalized density matrix $\hat \rho$, $\Tr\hat{\rho}=1$, the associated von Neumann entropy is
\begin{align}
S&\equiv -\Tr\hat{\rho}\ln \hat\rho\ .\label{vonNeumannEntropyDefinition}
\end{align}
The trick begins by re-expressing the logarithm as:
\begin{align}
S=-\lim_{n\to 1} \partial_{n}\left [\ln \left (\Tr\rho^{n}\right )-n\ln\left ( \Tr\rho\right )\right ]\ .\label{ReplicaTrickExpression}
\end{align}
In \eqref{ReplicaTrickExpression}, $\rho$ is an arbitrarily normalized density matrix related to $\hat{\rho}$ by  a constant factor, $\rho=\Tr(\rho)\hat\rho$.

The second part of the trick is the interpretation of \eqref{ReplicaTrickExpression} in terms of Euclidean time evolution and path integrals. Let $H(\tau)$ be some Hamiltonian describing our system of interest which is periodic\footnote{We review the trick in the general scenario where $H(\tau)$ has a periodic shift symmetry in $\tau$, but in all our cases of interest this is enhanced to a full, continuous $U(1)$ symmetry in the language of \cite{Lewkowycz:2013nqa}.} in imaginary time with $\tau\sim \tau+\beta$. We can use $H(\tau)$  to build the (non-normalized) density matrix
\begin{align}
 \rho=P \exp\left (-\int^{\tau_{2}}_{\tau_{1}}\rd\tau'\, H(\tau')\right )\label{DensityMatrixEuclidean}\ ,
 \end{align}
 where $P$ is the Euclidean time ordering operator, following the notation of \cite{Lewkowycz:2013nqa}.
When taking the trace of $\rho$, we adjust the integration limits to cover a single period
  \begin{align}
   \Tr\rho=\Tr P \exp\left (-\int_{0}^{\beta}\rd \tau'\, H(\tau')\right )\ .\label{DensityMatrixEuclideanTrace}
   \end{align} 
   The above describes a system whose dynamics are periodic in $\tau$ under shifts of $\tau\to \tau+\beta$ and which lives on a space whose Euclidean time coordinate also varies over this interval, $0\le \tau\le \beta$.

\begin{figure}[h!] 
  \captionsetup{width=0.9\textwidth}
   \centering
     \includegraphics[width=4in]{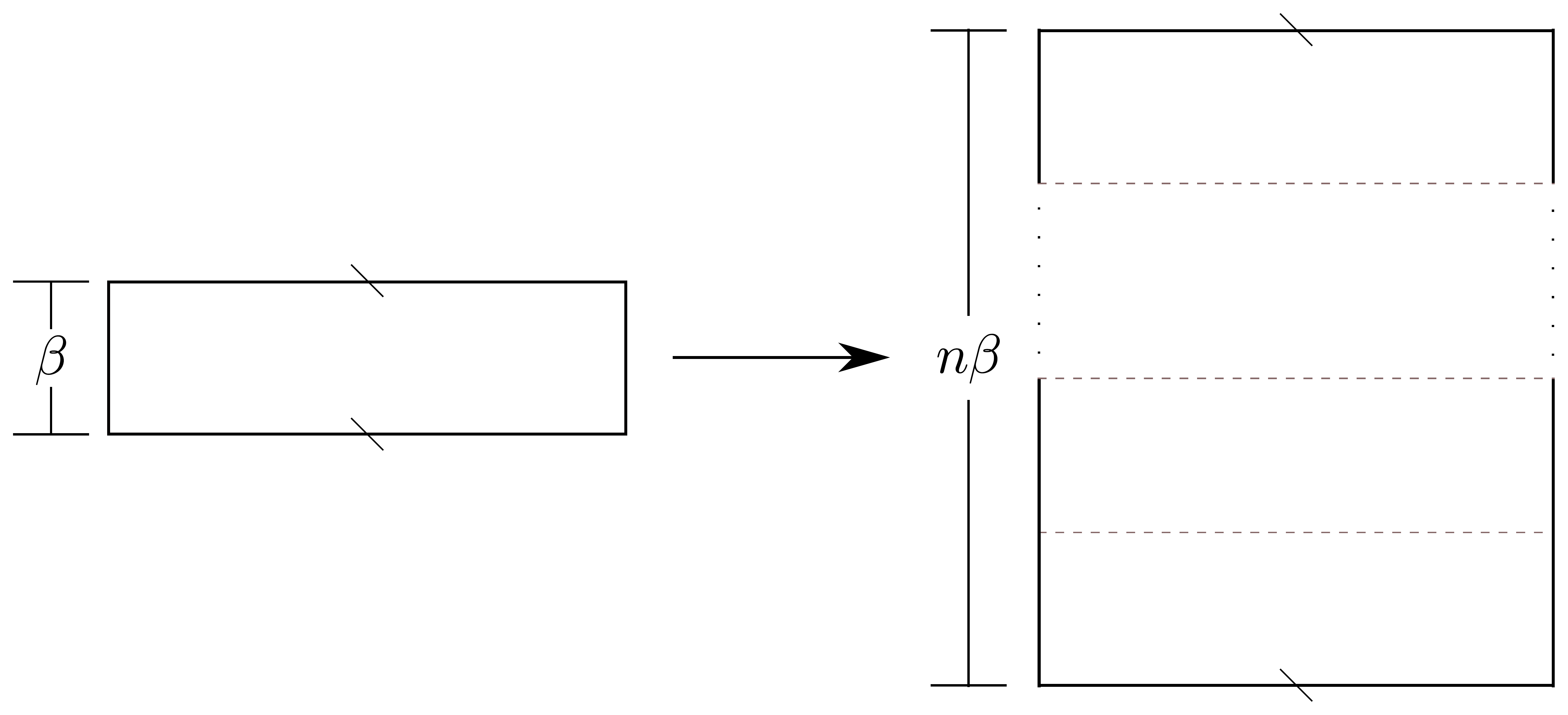}
   \caption{Setup of the replica trick. Time runs vertically, while horizontal slices represent spatial sections. The dynamics of the original (left) system are periodic under $\tau\to \tau+\beta$ and (anti)-periodic boundary conditions are also imposed, $\phi(0)=\pm\phi(\beta)$. The manifold is then extended to length $n\beta$, so that the fields now obey $\phi(0)=\pm\phi(n\beta)$, though the underlying Hamiltonian is unchanged and hence still periodic under $\tau \to \tau+\beta$, as indicated by the horizontal dashed lines.}
   \label{fig:ReplicaTrick}
\end{figure}

   We can then naturally interpret the $\Tr\rho^{n}$ term in \eqref{ReplicaTrickExpression} as the description of a system whose dynamics are the same as those above  (i.e.$\!$ it evolves with the \textit{same} Hamiltonian used in \eqref{DensityMatrixEuclidean} and \eqref{DensityMatrixEuclideanTrace}), but which now lives on a space whose Euclidean time coordinate now varies over the extended interval $0\le \tau\le n \beta$:
   \begin{align}
     \Tr\rho^{n}=P\exp\left (-\int_{0}^{n\beta}\rd \tau'\, H(\tau')\right )\ .
     \end{align}  The system is evolved for  $n-1$ extra cycles in $\tau$, resulting in $n$ ``replicas" of the original system, see Fig.$\!$ \ref{fig:ReplicaTrick}.

     Next, we can write the traces in terms of path integrals which can then be approximated by a saddle point calculation.  We have:
     \begin{align}
   \Tr\rho=\int\mathcal{D}\phi\, e^{-I^{(1)}_{\rm E}\left [\phi\right ]}\approx e^{-I^{(1)}_{\rm E}\left [\bar{\phi}^{(1)}\right ]}\ ,
     \end{align}
     where $\phi$ stands for all fields in the theory, $I^{(1)}_{E}\left [\phi\right ]$ is the Euclidean action living on a space where $\tau$ ranges over a single period $0\le \tau\le \beta$ (hence the $(1)$ superscript) and $\bar{\phi}^{(1)}$ is the solution which extremizes $I^{(1)}_{E}\left [\phi\right ]$.  As usual, the path integral is restricted to configurations obeying (anti-)periodic boundary conditions, $\phi(\tau)=\pm\phi(\tau+\beta)$, with the sign determined by the statistics of $\phi$.  
     
     Similarly, the trace of $\rho^{n}$ is given by
     \begin{align}
     \Tr\rho^{n}=\int\mathcal{D}\phi\, e^{-I^{(n)}_{\rm E}\left [\phi\right ]}\approx e^{-I^{(n)}_{\rm E}\left [\bar{\phi}^{(n)}\right ]}\ ,
     \end{align}
     where $I^{(n)}_{\rm E}\left [\phi\right ]$ is the Euclidean action living on a space where $\tau$ ranges over a $n$ periods $0\le \tau\le n\beta$ (hence the $(n)$ superscript) and $\bar{\phi}^{(n)}$ is the solution which extremizes $I^{(n)}_{E}\left [\phi\right ]$. Again, we only path integrate over periodic configurations, but now with an extended period $\phi(\tau)=\pm \phi(\tau+n\beta)$. The actions $I^{(1)}_{E}$ and $I^{(n)}_{E}$ share the same lagrangian and boundary conditions and only differ in their limits of integration. We are making the assumption here that, given some boundary conditions for the metric, we can find a smooth solution for arbitrary $n$. The entropy is therefore given by
     \begin{align}
     S&= \lim _{n\to 1}\partial_{n}\left (I^{(n)}_{\rm E}\left [\bar{\phi}^{(n)}\right ]-n I^{(1)}_{\rm E}\left [\bar{\phi}^{(1)}\right ]\right )\ .\label{EntropyInTermsOfActions}
     \end{align}

  In our cases of interest, the actions in \eqref{EntropyInTermsOfActions} will be gravitational and the metric configurations contain a horizon.  Then, when we study the Euclidean metrics which extremize the actions $I_{\rm E}^{(1)}$ and $I_{\rm E}^{(n)}$, the different Euclidean time intervals in the two cases enforce different near horizon geometries, due to the requirement of avoiding conical singularities. The replica metric $\bar g^{(n)}_{\mu\nu}$  is not equivalent to the familiar Euclideanized solution $\bar g^{(1)}_{\mu\nu}$.

     A final, particularly useful step is to analytically continue to $n=1+\epsilon$, with $\epsilon$ infinitesimal.  This allows us to avoid explicitly finding the $n\neq 1$ solution.  Rather than evaluating $I^{(1+\epsilon)}_{\rm E}$ on the true solution, $\bar{g}^{(1+\epsilon)}_{\mu\nu}$, we can evaluate on any off-shell solution $g^{\rm off}_{\mu\nu}$ which differs from the true one by an $\mathcal{O}(\epsilon)$ amount, since the error is second order in $\epsilon$:
    \begin{align}
 I_{\rm E}^{(1+\epsilon)}\left [\bar{g}_{\mu\nu}^{(1+\epsilon)}\right ]&= I_{\rm E}^{(1+\epsilon)}\left [g_{\mu\nu}^{\rm off}\right ]+\mathcal{O}(\epsilon^{2})\ .
    \end{align} The configuration $g^{\rm off}_{\mu\nu}$ doesn't solve the gravitational EOM and is simply chosen for convenience.  
    We only require that the true and off-shell metrics obey the same boundary conditions and that $g^{\rm off}_{\mu\nu}$ not contain conical singularities (under the identification $\tau\sim \tau+(1+\epsilon)\beta$).
    
    As $\bar{g}^{(1+\epsilon)}_{\mu\nu}$ and $\bar{g}^{(1)}_{\mu\nu}$ already differ only at $\mathcal{O}(\epsilon)$, we can simply build the off-shell metric from $\bar{g}^{(1)}_{\mu\nu}$ directly without ever considering the explicit form of $\bar{g}^{(1+\epsilon)}_{\mu\nu}$.   We only need to insert $\mathcal{O}(\epsilon)$  factors into $\bar{g}^{(1)}_{\mu\nu}$ to create $g_{\mu\nu}^{\rm off}$, subject to the requirements of the previous paragraph.    Therefore, we can take $g^{\rm off}_{\mu\nu}$ and $\bar g^{(1)}_{\mu\nu}$ to coincide everywhere except in an infinitesimal neighborhood near the horizon, where $g^{\rm off}$ must be altered to avoid a conical singularity.  
    
    This is one of the conceptual advantages of the replica trick: it allows us to isolate the entire non-trivial contribution to the entropy \eqref{EntropyInTermsOfActions} to the horizon, rather than have it arise, for instance, entirely from a boundary term at infinity, as it does in the standard Schwarzschild calculation \cite{Gibbons:1976ue}.  Relatedly, a significant practical advantage of the replica trick is that the contributions from any Gibbons-Hawking-York type boundary terms \cite{York:1972sj,Gibbons:1976ue} cancel out completely, since $g_{\mu\nu}^{\rm off}$ and $\bar g^{(1+\epsilon)}_{\mu\nu}$ obey the same boundary conditions at infinity \cite{Lewkowycz:2013nqa}. These properties\footnote{Wald's entropy formula \cite{Wald:1993nt,Iyer:1994ys} also shares these advantages.} allow us to treat asymptotically $dS$, $AdS$ and flat spacetimes in a single calculation.
    
    This somewhat abstract discussion is put into concrete use in the following section, where we derive a concrete expression for the entropy of spherically symmetric configurations for the action \eqref{PreciseAction}.  For derivations of more general entanglement entropy formulas for generic theories of gravity via the replica trick, see \cite{Dong:2013qoa,Camps:2013zua}.   We also present an explicit calculation of the $dS_{4}$ entropy via the replica trick in Appendix \ref{Appendix:dSReplicaTrick}.  

\subsubsection{Entropy: General Formulas and Limits}

In this section, we apply the replica trick to general, spherically symmetric solutions of \eqref{PreciseAction} and examine the resulting expressions in our regimes of interest.

Let $\bar g^{(1)}_{\mu\nu}$ be the Euclideanized version of the generic, spherically symmetric metric  \eqref{GenericSphericalMetric},
\begin{align}
\bar{g}^{(1)}_{\mu\nu}\rd x^{\mu}\rd x^{\nu}&=f(r)\rd \tau^{2}+\frac{1}{g(r)}\rd r^{2}+r^{2}\rd \Omega^{2}\ .\label{EuclideanizedGenericSphericalMetric}
\end{align}
We again assume that \eqref{EuclideanizedGenericSphericalMetric} has a horizon at $r=r_{h}$, resulting in a periodicity $\tau\sim \tau+\beta$ with $\beta=4\pi/\sqrt{f'(r_{h})g'(r_{h})}$ \eqref{GenericHawkingTemperature}.

The off-shell configuration, $g^{\rm off}_{\mu\nu}$, is then formed by taking \eqref{EuclideanizedGenericSphericalMetric} and inserting $\mathcal{O}(\epsilon)$ factors such that no conical singularity is introduced if we now identify $\tau\sim \tau+(1+\epsilon)\beta$.  This procedure is not unique.  If $\varepsilon(r)$ is a function equal to $\epsilon$ at the horizon, $\varepsilon(r_{h})=\epsilon$, but which quickly vanishes as $r$ is taken away from $r_{h}$ (so that any boundary conditions at infinity are preserved), then we can build a regular off shell metric with the desired periodicity by replacing\footnote{Similar procedures were used in \cite{Bhattacharyya:2013gra}.}, for instance, $f(r)\to f(r) \left (1+\varepsilon(r)\right )^{-2}$.  Alternatively, we could send $g(r)\to g(r)\left (1+\varepsilon(r)\right )^{-2}$ or work with the more general metric
\begin{align}
g^{\rm off}_{\mu\nu}\rd x^{\mu}\rd x^{\nu}&=\frac{f(r)}{\left (1+\varepsilon(r)\right )^{2-\alpha}}\rd \tau^{2}+\frac{\left (1+\varepsilon(r)\right )^{\alpha}}{g(r)}\rd r^{2}+r^{2}\rd \Omega^{2}\ , \label{GeneralOffShellEuclideanizedGenericSphericalMetric}
\end{align}
with $\alpha$ arbitrary.
As for the function $\varepsilon(r)$, we could, for instance, choose 
\begin{align}
\varepsilon(r)&=\epsilon \exp \left (-\frac{(r-r_{h})^{2}}{\ell^{2}}\right )\  \quad\ {\rm or} \quad
\varepsilon(r)=\epsilon \exp \left (-\kappa\frac{(r-r_{h})}{\ell}\right )\label{GaussianAndExponentialDamping}
\end{align}
where $\kappa=\pm 1$ with the sign determined by whether the horizon is cosmological ($-$ sign) or that of a black hole ($+$ sign).   In \eqref{GaussianAndExponentialDamping}, $\ell$ is a length scale which we take to zero at the end of the calculation so that $\bar g^{(1)}_{\mu\nu}$ and $g^{\rm off}_{\mu\nu}$ only differ from each other infinitesimally close to the horizon.

The computation is entirely straightforward, when using a computer to compute curvature tensors.  In some more detail, letting $\mathcal{L}_{\rm E}[g]$ be the Euclidean lagrangian, we calculate
\begin{align}
I_{\rm E}^{(1+\epsilon)}\left [\bar{g}^{(1+\epsilon)}\right ]&=\kappa\int_{0}^{(1+\epsilon)\beta}\rd \tau\int_{r_{h}}\rd r \int\rd^{2}\Omega\, \sqrt{g^{\rm off}}\, \mathcal{L}_{E}[g^{\rm off}]+\mathcal{O}(\epsilon^{2})\nn
I_{\rm E}^{(1)}\left [\bar{g}^{(1)}\right ]&=\kappa\int_{0}^{\beta}\rd \tau\int_{r_{h}}\rd r \int\rd^{2}\Omega\, \sqrt{\bar{g}^{(1)}}\, \mathcal{L}_{E}[\bar g^{(1)}]\label{ExplicitIs}
\end{align}
and derive the entropy via
    \begin{align}
     S&=\partial_{\epsilon}\left (I_{\rm E}^{(1+\epsilon)}\left [\bar{g}^{(1+\epsilon)}\right ]-(1+\epsilon) I_{\rm E}^{(1)}\left [\bar{g}^{(1)}\right ]\right )\Big|_{\epsilon=0}\ .\label{EntropyInTermsOfActionsAndEpsilon}
     \end{align}
     In \eqref{ExplicitIs}, the factors of $\kappa$ simply ensure that the radial integrals are taken in the correct direction, depending on whether the horizon is cosmological or that of a BH.  The lower limit on both radial integrals is determined by horizon position of $\bar{g}_{\mu\nu}^{(1)}$ and the upper limit depends on which type of horizon we are considering.
     
     The same entropy is found when using \eqref{GeneralOffShellEuclideanizedGenericSphericalMetric} with any value of $\alpha$ and either form of $\varepsilon(r)$ in \eqref{GaussianAndExponentialDamping}.  The answer for $g(r)\neq f(r)$ is, however, slightly unwieldy and relegated to Appendix \ref{Appendix:GeneralEntropyFormula} since, for the cases we're interested in and at the order we work to, we only require the result in the $g(r)\to f(r)$ limit, which is considerably cleaner. Writing the total entropy as the familiar GR area law result $S_{A}\equiv \frac{A}{4 G_{N}}= 8 \pi ^2 \left (r_{h}/l_{p}\right )^{2}$ and a correction, $S=S_{A}+\Delta S$, we find
     \begin{align}
     \Delta S
   &=-16 \pi ^2 \left(-(4a_1 +4a_{\rm GB})+  (8
   a_1 +2a_2) r_h f'+(2 a_1 +a_2)r_h ^2  f''\right)\nn
   &\quad -\frac{16 \pi ^2 f' }{m ^2 r_h }\left(8 b_1-   (4 b_1+2b_2)r_h f'+  (8 b_1+2b_2)
  r_h ^2 f''+\left (2 b_1
   +b_2\right ) r_h ^3
   f'''\right)\nn   
   &\quad +\frac{4 \pi ^2 }{m ^2 r_h ^2}\Big[ (16 c_2+48
   c_3+8c_5)-
    (192 c_3+32 c_5+8c_6)r_h  f'- (16 c_2+48
   c_3+8c_5) r_h ^2 f''\nn
   &\quad+ (16
   c_2+ 192 c_3+12c_4+48 c_5+12c_6)r_h ^2 f'^2  +
 (32 c_2+96 c_3+12 c_4+32 c_5+8
   c_6)r_h^{3}  f'f''\nn
   &\quad +(12 c_1+12 c_2+12
   c_3+4c_4+8 c_5+4c_6) r_h ^4 f''^2 \Big]\ .\label{GeneralEntropyResultForgttequalgrrinverse}
     \end{align}
    Above, primes are radial derivatives, arguments of $f$ are suppressed and all functions are evaluated at the horizon, $r=r_{h}$.  We've also replaced a $|f'|$ (arising from the periodicity $\beta\propto 1/|f'|$) with the equivalent expression $|f'|= \kappa f'$, a step which removes all factors of $\kappa$.
     
     Now we evaluate the entropy for the three cases emphasized in the previous sections, working to first order in EFT coefficients.  We write the total entropy as $S=\bar{S}+\delta S$ where $\bar{S}$ is the entropy of the horizon in pure GR and $\delta S$ is the correction which is first order in EFT coefficients.  We have:
\begin{itemize}
\item For small black holes $r_{s}\ll L_{(A)dS}$, the BH horizon entropy is
\begin{align}
\bar{S}&=8 \pi ^2 \left (\frac{r_{s}}{l_{p}}\right )^{2}+\frac{16}{3} \pi ^2 \Lambda  r_s ^4+\ldots\nn
\frac{\delta S}{\bar{S}}&=8 a_{\rm GB}\left (\frac{l_{p}}{r_{s}}\right )^{2}+4 c_1 \left (\frac{l_{p}}{mr_{s}^{2}}\right )^{2}+\ldots\ ,\label{EntropyShiftSmallRs}
\end{align}
where the ellipses contains terms suppressed by powers of $r_{s}/L_{(A)dS}\ll 1$.  
\item For large $SAdS$ black holes $r_{s}	\gg L_{AdS}$, the BH horizon entropy is
\begin{align}
\bar{S}&=8 \pi ^2  \left (\frac{L_{AdS}^{2}r_{s}}{l_{p}^{3}}\right )^{2/3}-\frac{16}{3} \pi ^2 \left (\frac{L_{AdS}}{l_{p}}\right )^{2} +\ldots\nn
\frac{\delta S}{\bar{S}}&=-(48 a_1+12 a_2)\left (\frac{l_{p}}{L_{AdS}}\right )^{2}+\ldots\nn
&\quad +(-28 c_1+16 c_2+672 c_3+42 c_4+168 c_5+42 c_6)\left (\frac{l_{p}}{mL_{AdS}^{2}}\right )^{2}+\ldots\ ,\label{EntropyShiftLargeRs}
\end{align}
where the ellipses contains terms suppressed\footnote{For instance, the Gauss-Bonnet term's contribution is $\delta S/\bar{S}\supset 8 a_{\rm GB}\left (\frac{l_{p}}{L_{AdS}}\right )^{2}\left (\frac{L_{AdS}}{r_{s}}\right )^{2/3}$.} by powers of $L_{AdS}/r_{s}\ll 1$. 
\item For small $SdS$ black holes, $r_{s}\ll L_{dS}$, the cosmological horizon entropy is
\begin{align}
\bar{S}&=8 \pi ^2 \left (\frac{L_{dS}}{l_{p}}\right )^{2}-8 \pi ^2\left (\frac{L_{dS} r_s}{l_{p}^{2}}\right ) +\ldots\nn
\frac{\delta S}{\bar{S}}&=(48 a_1+12 a_2+8 a_{\rm GB})\left (\frac{l_{p}}{L_{dS}}\right )^{2}\nn
&\quad +(16 c_1+96 c_2+576 c_3+36 c_4+144 c_5+36 c_6)\left (\frac{l_{p}}{mL_{dS}^{2}}\right )^{2}+\ldots\label{EntropyShiftCos}
\end{align}
where the ellipses contains terms suppressed by powers of $r_{s}/L_{dS}\ll 1$.  
\end{itemize}
Above, we've traded mass scales for length scales when convenient and given fractional corrections to emphasize the perturbative nature of the calculation: $\delta S/\bar{S}\ll 1$ when the EFT description is valid \eqref{EFTValidityConditions}, $L_{(A)dS}, r_{s}\gg m^{-1}\gg l_{p}$.  Again, Euclidean $SdS$ calculations are conceptually confusing for the reasons discussed at the end of Sec.$\!$ \ref{Sec:HawkingRadiation}.  For $dS$ asymptotics, the periodicities associated to the black hole and cosmological horizons (\eqref{THShiftLargeRs} and \eqref{THShiftCos}, respectively) were used in the calculations \eqref{EntropyShiftSmallRs} and \eqref{EntropyShiftCos}, respectively.

The above results require a few comments.  First, in each result \eqref{EntropyShiftLargeRs}, \eqref{EntropyShiftLargeRs} and \eqref{EntropyShiftCos}, the correction $\delta S$ is dominated by the $a_{i}$ coefficients, assuming the $a_{i}$'s are at least as large as their natural $\mathcal{O}(1)$ value.  The contributions from the $c_{i}$'s are suppressed by factors of $(m r_{s})^{-2}$ or $(mL_{(A)dS})^{-2}$ and the contributions from $b_{i}$'s (not included above) are smaller yet.  However, because the value of the $a_{i}$'s is ambiguous, we will mostly concentrate on the $c_{i}$ terms which represent the leading corrections which are unambiguously determined by the theory's matter content.

Next, the six-derivative operators contribute to $\delta S$ both through shifts in the  area law term (since the $c_{i}$ operators shift the horizon location) and through their non-area law corrections to the general form of the entropy. These two effects are the same order of magnitude. For instance, the $\sim R^{3}$ operators shift the horizon of a Schwarzschild black hole by an amount $\delta r_{h}\sim r_{s}\left (\frac{l_{p}^{2}}{m^{2}r_{s}^{4}}\right )$ which changes the area law term by $\delta S_{A}\sim  r_{s}\delta r_{h}/l_{p}^{2}\sim \left (mr_{s}\right)^{-2}$ and an operator $\sim R^{3}/m^{2}$ also contributes $\delta S\sim \left (mr_{s}\right )^{-2}$ to the entropy, by dimensional analysis.

Finally, we should compare the results obtained via the replica trick to others' results obtained through different means.  Reference \cite{Lu:1993sq} considered the effect of $\sim R^{3}$ operators on Schwarzschild black holes in $d=4$ and determined the correction to the entropy both by evaluating the free energy of the solution and by using the thermodynamic relation $\delta E=T\delta S$. After using the identities \eqref{DimensionDependentIdentitiesExplicit} to map their action to ours, we find perfect agreement.  Next, \cite{Matyjasek:2006fq} generalized the results of \cite{Lu:1993sq} to arbitrary $d$ while also including the $\sim R^{2}$ operators in $I_{\rm EFT}^{\partial^{2}}$ \eqref{PreciseAction}.  They calculated the entropy using Wald's formula and their results agree with ours in the $d\to 4$ limit.  In particular, they find the same contribution from the Gauss-Bonnet term\footnote{See also, for instance, \cite{Liko:2007vi,Sarkar:2010xp} who find the same result and \cite{Chatterjee:2013daa} who differ by a factor of two.}, $S\supset 64\pi^{2}a_{\rm GB}$, which is common to every horizon due to the topological nature of $\mathcal{L}_{\rm GB}$.

 \subsection{The First Law\label{Sec:FirstLaw}}

 Given the corrected expressions for the Hawking temperature and entropy, we can explore the first law $\delta E=T \delta S$ in the cases of both asymptotically flat and large $AdS$ black holes.
 
 Lu and Wise \cite{Lu:1993sq} previously found the corrections to Schwarzschild black holes from higher derivative operators and confirmed that the first law holds.  As we agree with their results, it is no surprise that we also find the first law to be satisfied.  The corrected Hawking temperature and entropy are, to first order,
 \begin{align}
 T_{H}&=\frac{1}{4\pi r_{s}}+\frac{c_{1}}{\pi m^{2}M_{p}^{2}r_{s}^{2}}\nn
 S&=8\pi^{2}M_{p}^{2}r_{s}^{2}+64\pi^{2}a_{\rm GB}+\frac{32c_{1}\pi^{2}}{m^{2}r_{s}^{2}}\label{THandSSchwarzschild}\ .
 \end{align}
 Varying $r_{s}\to r_{s}+\delta r_{s}$, we find
 \begin{align}
 T_{H}\delta S&=4\pi M_{p}^{2}\delta r_{s}=\delta M=\delta E_{\rm ADM}\ ,
 \end{align}
 as $r_{s}\equiv M/(4\pi M_{p}^{2})$, and so the first law holds.  Above, we've noted that the energy of asymptotically flat black holes is not shifted from its GR value due to higher derivative corrections, $E_{\rm ADM}=M$, due to the reasoning in Footnote \ref{Foot:MisEFootnoteFlatSpace}.

 The situation is different for large $AdS$ black holes, however.  For instance, the $a_{i}$ operators do not alter the solution at first order in $d=4$ \eqref{VanishingContributionOfAOperatorsin4D} and hence they don't change the Hawking temperature of large $AdS$ black holes.  However, they \textit{do} change the entropy \eqref{EntropyShiftLargeRs} and hence they must also affect the $AdS$ (free) energy if $\delta E=T\delta S$ is to hold.
 
 Calculating $T\delta S$ from \eqref{THShiftLargeRs} and \eqref{EntropyShiftLargeRs}, we deduce that the corrected $AdS$ free energy expression is just a rescaling of the original result:
 \begin{align}
 E_{\rm AdS}&=M\times \left (1-\frac{(48a_{1}+12a_{2})}{M_{p}^{2}L_{\rm AdS}^{2}}+ \frac{\left (-24 c_{1}+48 c_{2}+864 c_{3}+54 c_{4}+216 c_{5}+54 c_{6}\right )}{m^{2}M_{\rm pl}^{2}L_{\rm AdS}^{4}}\ldots\right )\label{CorrectedAdSFreeEnergy}\ .
 \end{align}
 The $a_{i}$ terms represent the largest correction, but these coefficients are not primarily determined by the properties of the heavy matter fields in the theory.  Rather it's the coefficients of the (classically) irrelevant operators which are dictated by heavy matter and hence we retain the $c_{i}$'s contributions above, too.   
 
  The corrections due to $a_{i}$ terms in \eqref{CorrectedAdSFreeEnergy} agrees with \cite{Smolic:2013gz} who found the explicit boundary stress tensor for four-dimensional $AdS$ gravity with quadratic $\sim R^{2}$ operators added to the action.   
  It would be interesting to explicitly calculate the contribution of $\sim R^{3}$ operators to the holographic stress tensor \cite{Balasubramanian:2001nb} and confirm the contribution of the $c_{i}$'s in  \eqref{CorrectedAdSFreeEnergy}, but we leave this to future work.

\section{Effective Actions\label{Sec:EffectiveActions}}

In this section, we discuss the construction of the effective action which arises when heavy fields are integrated out and present the results of specific calculations.  The presentation is somewhat brief and schematic as the ideas and results are standard.

\subsection{Integrating Out Heavy Matter\label{Sec:EffectiveActionsIntegratingOut}}

Given some set of heavy matter fields, denoted by $\Phi$, described by an action $I_{m}[\Phi,g_{\mu\nu}]$, we can capture the small effects of these fields on low energy physics by simply integrating them out of the action.  In the context of gravitational physics, we can split the action into one part which depends on the heavy matter and another which is purely gravitational
\begin{align}
I=I_{\rm grav}[g_{\mu\nu}]+I_{m}[\Phi,g_{\mu\nu}]\ .\label{ISplitIntoIgravAndIm}
\end{align}
At low energies, it is then appropriate to work with the EFT generated by integrating out $\Phi$:
\begin{align}
\exp \left (iI_{\rm EFT}[g_{\mu\nu}]\right )&\equiv \int\mathcal{D}\Phi\, \exp \left (iI_{\rm grav}[g_{\mu\nu}]+iI_{m}[\Phi,g_{\mu\nu}]\right )\ .\label{SEFTPathIntegralDefinition}
\end{align}

There are two practical issues in calculating \eqref{SEFTPathIntegralDefinition}: the answer cannot be computed exactly and it is also divergent.  Both issues are easily dealt with.

  First, the path integral admits an expansion in powers of $R_{\mu\nu\rho\sigma}/m^{2}$ and $\nabla^{2}/m^{2}$, where $m$ is the mass of the heavy particle; see Appendix \ref{Appendix:RulesForMatrixElements}.  Operators generated via the path integral only induce small effects at low energies and one can consistently truncate the action at a given order in this derivative expansion. Working to six-derivative order is sufficient for the interests of this paper, for example.
  
  Second, divergences are treated in the standard way by introducing counterterms and renormalized quantities.  For instance, integrating a generic field $\Phi$ of mass $m$ will generate a divergent term $\propto R$.  Using dimensional regularization with $d=4-\delta$, this divergence will enter the effective action as
  \begin{align}
  \exp \left (iI_{\rm EFT}[g_{\mu\nu}]\right )=\exp \left (iI_{\rm grav}[g_{\mu\nu}]+i\int\rd^{4}x\sqrt{-g}\, \frac{bm^{2}}{(4\pi)^{2}}\left (\frac{1}{\delta}-\frac{1}{2}\ln m^{2}/\mu ^{2}\right )R+\ldots\right )
  \end{align}
  where $b$ is some $\mathcal{O}(1)$ number and $\mu$ is the arbitrary RG scale.  Within $I_{\rm grav}$, there's a corresponding ``bare" $\sim R$ term 
  \begin{align}
  I_{\rm grav}[g_{\mu\nu}]=\frac{M_{(b)}^{2}}{2}\int\rd^{4}x\sqrt{-g}\, R+\ldots\ ,
  \end{align}
  where $M_{(b)}$ is the bare Planck mass.  The divergence is dealt with by defining
  \begin{align}
  M_{(b)}^{2}&=M_{p}^{2}-\frac{1}{\delta}\frac{2bm^{2}}{(4\pi)^{2}}\label{RenormalizedMp}
  \end{align}
  which also defines the renormalized Planck mass, $M_{p}$, which is what appears in physical expressions.  
  
  This procedure also determines the one-loop beta functions of the theory.  Focusing on the Planck mass still, the effective action contains the terms
  \begin{align}
  I_{\rm EFT}[g_{\mu\nu}]=\int\rd^{4}x\sqrt{-g}\, \left (\frac{M_{p}^{2}}{2}-\frac{bm^{2}}{2(4\pi)^{2}}\ln m^{2}/\mu ^{2}\right )R+\ldots\ .
  \end{align}
  Above, $\mu$ is an entirely arbitrary scale whose value cannot affect physical results.  This means that $M_{p}^{2}$ must also depend on $\mu$ and its dependence is determined by demanding that $I_{\rm EFT}$ be $\mu$-independent\footnote{This is sufficient for the present applications, but more generally one also has to account for wavefunction renormalization and the procedure for finding beta functions and anomalous dimensions is (only slightly) more complicated.   See \cite{GeorgiWeak} for the more general treatment.}
  \begin{align}
\frac{\rd M_{p}^{2}}{\rd \ln \mu}\equiv   \beta(M_{p}^{2})=-\frac{2bm^{2}}{(4\pi)^{2}}\label{MpBetaFunctionExample}\ .
  \end{align}
  This beta function is only valid for energy scales $\mu\gg m$.  At low energies heavy matter decouples and we can take $\beta(M_{p}^{2})$ to vanish for $\mu \ll m$  \cite{Manohar:1996cq} so that the behavior of $M_{p}^{2}$ is roughly
  \begin{align}
  M_{p}^{2}(\mu)&=\begin{cases}M_{p}^{2}(m)-\frac{bm^{2}}{(4\pi)^{2}}\ln \mu ^{2}/m^{2} & \mu \gg m\\
  M_{p}^{2}(m) & \mu \ll m
  \end{cases}\label{MpBehaviorRough}\ ,
  \end{align}
  where any running from other possible diagrams was ignored and $M_{p}^{2}(m)$ is the value of $M_{p}^{2}$ at $\mu\approx m$.  In four dimensions, there are also divergences corresponding to the  Cosmological Constant and the $\sim R^{2}$ operators whose corresponding beta functions are determined similarly to the above.  In each of these cases, the renormalized value of the coupling of interest are strongly determined by UV physics, not just by properties of the heavy matter fields.  As discussed before, we are tuning bare parameters so that the GR and EFT actions share the IR same values of $M_{p}$ and $\Lambda$.

  Most importantly, there are finite contributions from the path integral which are unamgibuously determined by the properties of the matter field.  Heavy fields generate operators of the form $\sim R^{n}/m^{2(n-2)}$, $n\ge 3$, for instance, which are suppressed by appropriate powers of $m$ and come with finite numerical factors.  Such irrelevant operators are then added to those which may have already existed in the gravitational action $I_{\rm grav}$ \eqref{ISplitIntoIgravAndIm}.  However, it is natural to expect that these latter operators are negligible, as compared to those those we've just generated.  This is because the irrelevant operators already included in $I_{\rm grav}$ should be suppressed by a much larger energy scale, some $\Lambda_{\rm UV}\gg m$, corresponding to even heavier physics which has already been integrated out. These operators therefore look like $\sim R^{n}/\Lambda_{\rm UV}^{2(n-2)}$ and are overtaken by the new $\sim R^{n}/m^{2(n-2)}$ terms.  Thus, in what follows we will calculate the leading, irrelevant operators which arise from different forms of matter and take them to be the dominant terms in the expansion.

When calculating the leading results of the path integral,  we work with flat space propagators for all of the matter fields.  This is appropriate for the scenarios under consideration, as we're taking the Compton wavelength of the heavy fields, $m^{-1}$, to be much smaller than all other curvature length scales in the problem.  In this regime the matter feels as though it is in flat space.  

Our treatment also misses some non-perturbative effects which aren't captured by effective actions, such as the exponentially suppressed $\sim e^{-m/H}$ signatures of particle production in $dS$ recently studied in \cite{Arkani-Hamed:2015bza,Lee:2016vti}.  Such contributions are expected to be negligible, since their characteristic exponential suppression is much stronger than the power law suppression arising from the higher derivative operators under study.
  
  \subsection{Application to Scalars, Fermions and Vectors}
  
  In this section, we quickly discuss the calculation of the path integral for free, minimally coupled scalars, Dirac fermions and vector fields, each of mass $m$.  
  Since we're considering only non-self-interacting fields, the path integrals in each case yield relatively simple functional determinants.  
  
  The scalar action is
  \begin{align}
  I_{m}[\phi]&=\frac{1}{2}\int\rd^{4}x\sqrt{-g}\, \phi\left(\square-m^{2}\right )\phi
  \end{align}
  and integrating out $\phi$ yields the familiar result\footnote{In the logarithm, we've dropped a $\sqrt{-g}$ factor, as its contribution is vanishing in dimensional regularization: $\Tr\ln \sqrt{-g}(\square-m^{2})=\Tr\ln \sqrt{-g}+\Tr\ln (\square-m^{2})$ and $\Tr\ln \sqrt{-g}\propto\delta^{d}(0)$, which vanishes in this regularization.}
  \begin{align}
  \int\mathcal{D}\phi\, \exp  iI[\phi]=\exp\left (-\frac{1}{2}\Tr\ln \left (\square-m^{2}\right )\right )\ .
  \end{align}
  
  The fermion action is
  \begin{align}
  I_{m}[\psi,\bar{\psi}]&=\int\rd^{4}x\det (e)\, \bar{\psi}\left (i\slashed \nabla-m\right )\psi	
  \end{align}
  where $e_{\mu}{}^{a}$ is the vielbein and $\slashed \nabla$ the spinor covariant derivative written in terms of the spin connection $\omega_{\mu bc}$:
  \begin{align}
  \slashed \nabla\equiv \gamma^{a}e^{\mu}{}_{a}\left (\partial_{\mu}+\frac{1}{4}\gamma^{b}\gamma^{c}\omega_{\mu bc}\right )\quad , \quad\omega_{\mu bc}\equiv e_{\beta b}e^{\sigma}{}_{c}\Gamma^{\beta}_{\sigma \mu}-e^{\sigma}{}_{c}	\partial_{\mu}e_{\sigma b}\ , 
    \end{align}
    with $\gamma^{a}$ the usual gamma matrices.  The resulting functional determinant is:
    \begin{align}
    \int\mathcal{D}\psi\mathcal{D}\bar{\psi}\, \exp iI_{m}[\psi,\bar{\psi}]&=\exp \Tr\ln (i\slashed\nabla-m)\ .
    \end{align}
    There's no factor of $-1/2$ in front of the trace due to the complex, grassmannian nature of the fields.
    
    Finally, we treat the massive vector field. We study these in isolation, though they should of course be UV completed via some Higgs field which we're taking to be much heavier than the vector.  We perform the calculation in two different ways.  The starting point is the action
    \begin{align}
    I_{m}[A_{\mu}]&=\int\rd^{4}x\sqrt{-g}\, \left (-\frac{1}{4}F_{\mu\nu}^{2}-\frac{m^{2}}{2}A_{\mu}^{2}\right ) \ .\label{NaiveVectorAction}
    \end{align}
    The path integral can be performed directly to yield
    \begin{align}
    \int\mathcal{D}	A_{\mu}\, \exp iI_{m}[A_{\mu}]&=\exp\left  ( -\frac{1}{2}\Tr\ln \left (\square \delta^{\mu}_{\nu}-\nabla_{\nu}\nabla^{\mu}-m^{2}\delta^{\mu}_{\nu}\right )\right )\ ,\label{NaiveVectorFunctionalDeterminant}
    \end{align}
    where the covariant derivatives are to be treated as acting \textit{only} on the $\mu$ index of $\delta^{\mu}_{\nu}$.  The functional determinant can then be evaluated perturbatively, but there's a mild technical annoyance: the massive vector propagator is $D_{\mu\nu}(p)\propto \left (\eta_{\mu\nu}+p_{\mu}p_{\nu}/m^{2}\right )/(p^{2}+m^{2})$ which is undesirable for two reasons.  First, it (roughly speaking) goes to a constant in the UV, rather than falling off as $D_{\mu\nu}\sim p^{-2}$ and, second, the pesky $\sim p_{\mu}p_{\nu}$ factors result in rather complicated loop integrals.
    
    A more clever treatment of the massive vector borrows standard techniques from studies of spontaneously broken gauge theories.  We use the Stuckelberg trick \cite{Stueckelberg:1957zz}, nicely reviewed in \cite{Hinterbichler:2011tt}, and consider the action we get by replacing $A_{\mu}\to A_{\mu}+\frac{1}{m}\partial_{\mu}\pi$ in \eqref{NaiveVectorAction}:
    \begin{align}
    I_{m}[A_{\mu},\pi]&=\int\rd^{4}x\sqrt{-g}\,\left ( -\frac{1}{4}F_{\mu\nu}^{2}-\frac{m^{2}}{2}A_{\mu}^{2}-\frac{1}{2}(\partial_{\mu}\pi)^{2}-mA^{\mu}\partial_{\mu}\pi\right )\ .\label{StuckelbergVectorAction}
    \end{align}
    Though we've introduced a new scalar field $\pi(x)$, the theories \eqref{StuckelbergVectorAction} and \eqref{NaiveVectorAction} are equivalent, since \eqref{StuckelbergVectorAction} has also gained the gauge symmetry
    \begin{align}
    A_{\mu}\to A_{\mu}+\partial_{\mu}\chi \ , \quad \pi \to \pi -m\chi\ .
    \end{align}
    Essentially, we've reintroduced the Goldstone mode $\pi(x)$ which would appear after Higgsing.
    
    The technical advantage in using this trick is that it allows us to remove the $\sim p_{\mu}p_{\nu}$ terms in the vector propagator, greatly simplifying the calculation.  We accomplish this by gauge fixing and going to Feynman-'t Hooft gauge where propagators take on the much more manageable $D(p)\sim \eta_{\mu\nu}/p^{2}$ form.  Following the usual Faddeev-Popov procedure and using the gauge fixing function $G=\partial^{\mu}A_{\mu}+m \pi$ results in the action
    \begin{align}
    I_{m}[A_{\mu},\pi,c,\bar{c}]&=\int\rd^{4}x\sqrt{-g}\left (\frac{1}{2}A^{\mu}\square A_{\mu}-\frac{m^{2}}{2}A_{\mu}^{2}+\frac{1}{2}\pi\left (\square-m^{2}\right )\pi+\bar{c}\left (\square-m^{2}\right )c\right )\ ,\label{GaugeFixedStuckelbergedActionVectors}
    \end{align}
    where $c,\bar{c}$ are the ghost fields.  The path integral yields
    \begin{align}
    \int\mathcal{D}\bar{c}\mathcal{D}c\mathcal{D}\pi\mathcal{D}A_{\mu}\exp iI_{m}[A_{\mu},\pi,c,\bar{c}] &=\exp \left (-\frac{1}{2}\Tr\ln \left (\square\delta^{\mu}_{\nu}-m^{2}\delta^{\mu}_{\nu}\right )+\frac{1}{2}\Tr\ln \left (\square-m^{2}\right )\right )\ , \label{StuckelbergedVectorPathIntegral}
    \end{align}
    which is far simpler to calculate than \eqref{NaiveVectorFunctionalDeterminant}.    Again, $\square$ in the first term of \eqref{StuckelbergedVectorPathIntegral} only acts on the $\mu$ index of $\delta^{\mu}_{\nu}$ and the overall $+1/2$ coefficient in the second term came from the combination of the real scalar field and complex, grassmannian scalar ghost fields in \eqref{GaugeFixedStuckelbergedActionVectors}.  
    
    All of the traces were calculated perturbatively to sixth order in derivatives using the rules in Appendix \ref{Appendix:RulesForMatrixElements} and extensive use of the $xAct$ package \cite{xAct}. In particular, both \eqref{NaiveVectorFunctionalDeterminant} and \eqref{StuckelbergedVectorPathIntegral} yielded the same answer. The results for the beta functions are\footnote{In previous versions of this manuscript the numerical values of $\beta(a_{1})$ and $\beta(a_{2})$ were incorrect, as the results of the one-loop calculation were mistakenly matched to $\mathcal{L}_{\rm EFT}^{\partial^{4}}=a_{1}R^{2}+a_{2}R_{\mu\nu}^{2}+a_{\rm GB}R_{\mu\nu\rho\sigma}^{2}$, rather than $\mathcal{L}_{\rm EFT}^{\partial^{4}}=a_{1}R^{2}+a_{2}R_{\mu\nu}^{2}+a_{\rm GB}\mathcal{L}_{\rm GB}$ with $\mathcal{L}_{\rm GB}$ as in \eqref{GaussBonnet}.  This error had no effect on our conclusions.}
    \begin{align}
    \rowcolors{2}{gray!5}{white}
    \begin{tabular}{ | c || c | c | c | c | c | c | c | c | c | c | c | c | c |}
    \hline Matter  & $  \beta(\Lambda) $ & $  \beta(M_{p}^{2}) $ & $  \beta(a_{1})$ & $  \beta(a_{2}) $ & $\beta(a_{\rm GB})$	\\
     \hline Scalar & $\frac{m^{4}}{ 2 (4\pi)^{2}}$   & $\frac{m^{2}}{ 3 (4\pi)^{2}}$   & $-\frac{1}{ 120 (4\pi)^{2}}$   & $-\frac{1}{ 60 (4\pi)^{2}}$   & $-\frac{1}{ 180 (4\pi)^{2}}$  \\
     \hline	 Fermion & $-\frac{2m^{4}}{  (4\pi)^{2}}$   & $\frac{2}{3}\frac{m^{2}}{  (4\pi)^{2}}$   & $\frac{1}{ 30 (4\pi)^{2}}$   & $-\frac{1}{ 10 (4\pi)^{2}}$   & $-\frac{7}{ 360 (4\pi)^{2}}$  \\
       \hline Vector & $\frac{3}{2}\frac{m^{4}}{  (4\pi)^{2}}$   & $-\frac{m^{2}}{  (4\pi)^{2}}$   & $\frac{7}{ 120 (4\pi)^{2}}$   & $-\frac{13}{ 60 (4\pi)^{2}}$   & $\frac{1}{ 15 (4\pi)^{2}}$  \\
     \hline	
    \end{tabular}\ ,\label{EFTBetaFunctions}
    \end{align}
    while the finite $b_{i}$'s are found to be
     \begin{align}
      \rowcolors{2}{gray!5}{white}
    \begin{tabular}{ | c || c | c |}
    \hline Matter & $  b_{1} $ & $  b_{2}$ \\
     \hline Scalar &  $\frac{1}{672(4\pi)^{2}}$ &  $\frac{1}{1680(4\pi)^{2}}$  \\
      \hline Fermion &  $-\frac{1}{560(4\pi)^{2}}$ &  $\frac{1}{168(4\pi)^{2}}$ \\
       \hline Vector &  $-\frac{9}{1120(4\pi)^{2}}$ &  $\frac{3}{112(4\pi)^{2}}$\\ \hline
    \end{tabular}\label{EFTbiValues}
    \end{align}
    and, finally, the finite $c_{i}$'s are given by
     \begin{align}
      \rowcolors{2}{gray!5}{white}
    \begin{tabular}{ | c ||  c | c | c | c | c | c | c | c | c | c | c |}
    \hline Matter & $  c_{1} $ & $  c_{2} $ & $  c_{3} $ & $  c_{4} $ & $  c_{5} $ & $  c_{6}  $	\\
     \hline Scalar &  $\frac{1}{30240(4\pi)^{2}}$ &  $\frac{1}{1680(4\pi)^{2}}$ &  $\frac{1}{1680(4\pi)^{2}}$ &  $\frac{1}{1512(4\pi)^{2}}$ &  $-\frac{1}{504(4\pi)^{2}}$ &  $\frac{1}{630(4\pi)^{2}}$ \\
      \hline Fermion &  $-\frac{1}{7560(4\pi)^{2}}$ &  $\frac{1}{2520(4\pi)^{2}}$ &  $\frac{1}{560(4\pi)^{2}}$ &  $\frac{43}{7560(4\pi)^{2}}$ &  $-\frac{13}{1008(4\pi)^{2}}$ &  $\frac{1}{63(4\pi)^{2}}$ \\
       \hline Vector &  $\frac{1}{10080(4\pi)^{2}}$ &  $-\frac{1}{1008(4\pi)^{2}}$ &  $\frac{1}{2520(4\pi)^{2}}$ &  $-\frac{37}{2520(4\pi)^{2}}$ &  $-\frac{11}{1260(4\pi)^{2}}$ &  $\frac{11}{252(4\pi)^{2}}$\\ \hline
    \end{tabular}\ .\label{EFTciValues}
    \end{align}

  We can compare the finite $b_{i}$ and $c_{i}$ coefficients to Table 1 of \cite{Avramidi:1986mj}, for instance, which performs the same calculation using covariant, proper time techniques.  We find agreement after accounting for dimension-dependent identities \cite{Edgar:2001vv}, up to the fact that our result for the Dirac fermions is exactly twice as large as theirs.  It seems that the factor of two\footnote{Reference \cite{Matyjasek:2006nu} studies non-minimally coupled matter fields and their EFT coefficients very nearly coincide with ours in the minimally coupled limit (after taking into account conventions and identities).  The only coefficient which causes any discrepancy is their $\alpha_{6}^{(1/2)}$ in Table I of \cite{Matyjasek:2006nu} which reads $-25/376$, but which we believe should read $-25/378$.  For instance, all other $\alpha_{i}^{(1/2)}$'s are exactly twice as large as the corresponding $c_{i}$'s in Table I of \cite{Avramidi:1986mj} and this would also be true of $\alpha_{6}^{(1/2)}$ if it took on the value $-25/378$.} which appears in front of the functional determinant when integrating out complex fields may have been lost in \cite{Avramidi:1986mj}.

\section{Results for Asymptotically Flat, $dS$ and $AdS$ Spacetimes\label{Sec:AllAsymptoticPossibilities}}

In the following sections, we collect the results for asymptotically flat, $dS$ and $AdS$ spacetimes evaluating our general formulas on the specific EFT coefficients \eqref{EFTbiValues} and \eqref{EFTciValues} derived in the previous section.  We present the results for the case where we've integrated out a single scalar, fermion or vector of mass $m$.  The extension to multiple species is trivial.

\subsection{Asymptotically Flat Black Holes\label{Sec:AsymptoticallyFlatBHs}}

Start with the asymptotically flat case.

Up to higher order corrections, the horizon of asymptotically flat black holes now occurs at:
\begin{align}
r_{h}&=r_s+\frac{1}{241920\pi^{2}}\left ( \frac{l_{p}^{2}}{m^{2}r_{s}^{3}}\right )\times\begin{cases}
-113 & {\rm scalar} \\
-52& {\rm fermion} \\
+165& {\rm vector} 
\end{cases} \label{HorizonsSmallBH}\ 
\end{align} and the corrected Hawking temperature is given by
\begin{align}
T_{H}&=\frac{1}{4 \pi 
   r_s }+\frac{1}{483840\pi^{3}}\left (\frac{l_{p}^{2}}{m^{2}r_{s}^{5}}\right ) \times\begin{cases}
+1 & {\rm scalar} \\
-4& {\rm fermion} \\
+3& {\rm vector} 
\end{cases}\label{THSmallBH}
\end{align}
where the first terms in \eqref{HorizonsSmallBH} and \eqref{THSmallBH} are GR answers and the second terms are the leading correction from the EFT operators.  
We discuss the signs appearing above in Sec.$\!$ \ref{Sec:Interpretation}.

The corrected entropy is
\begin{align}
S&=8 \pi ^2
   \left (\frac{r_{s}}{l_{p}}\right)^{2}+ 64 \pi ^2 a_{\rm GB}+\frac{1}{15120}\left (\frac{1}{mr_{s}}\right )^{2} \times\begin{cases}
+1 & {\rm scalar} \\
-4 & {\rm fermion} \\
+3& {\rm vector} 
\end{cases} \label{EntropySmallBH}
\end{align}
where the first term in \eqref{EntropySmallBH} is the GR answer, the second term is the leading correction from the $a_{i}$ EFT operators and the third term is the leading correction from the six-derivative, $c_{i}$ operators (the $b_i$'s don't contribute at leading order).  
The result \eqref{EntropySmallBH} agrees with the findings of \cite{Matyjasek:2006nu} in the scalar and vector cases, but not for fermions where our answer is precisely twice as large as theirs; see their (35)-(37) (after some translation, in the appropriate limit). 

Note that shifts to the the Hawking temperature and black hole entropy, but not the horizon radius, vanish completely if the matter content is SUSY-like\footnote{We thank Guilherme Pimentel for discussions on this point.} in the sense that all species have the same mass and that they collectively induce no running for the cosmological constant.  That is, the running of $\Lambda$ from heavy matter is
\begin{align}
\beta(\Lambda)=\frac{m^{4}}{2(4\pi)^{2}}\left (N_{0}-4N_{1/2}+3N_{1}\right )
\end{align}
where $N_{s}$ is the number of spin-$s$ particles of mass $m$ in the spectrum, and the corrections are also proportional to this same factor, $\delta T_{H}\propto	\delta S\propto	\left (N_{0}-4N_{1/2}+3N_{1}\right )/m^{2} $.  It would be interesting to see if this property persists at higher orders.

\subsection{Large, Asymptotically $AdS$ Black Holes\label{Sec:AsymptoticallyAdSBHs}}

Next, we study asymptotically $AdS$ black holes.  We focus on large black holes, $r_{s}\gg L_{AdS}$, as the changes to small $AdS$ black holes are essentially the same as in Sec.$\!$ \ref{Sec:AsymptoticallyFlatBHs} and, treated thermodynamically, small black holes never dominate the ensemble anyway.

Up to higher order corrections, the horizon for a large BH in $AdS $ occurs at
\begin{align}
r_{h}&= \left (r_{s}L_{AdS}^{2}\right )^{1/3}+\frac{1}{241920\pi^{2}}\left (\frac{r_{s}}{L_{AdS}}\right )^{1/3}\left (\frac{l_{p}^{2}}{m^{2}L_{AdS}^{3}}\right )\times\begin{cases}
-607& {\rm scalar} \\
-8& {\rm fermion} \\
+363& {\rm vector} 
\end{cases} \label{HorizonsLargeBH}
\end{align} and the corrected Hawking temperature is given by
\begin{align}
T_{H}&=\frac{3 }{4 \pi  L_{AdS}}\left (\frac{r_{s}}{L_{AdS}}\right )^{1/3}+\frac{1}{161280\pi^{3}}\left (\frac{r_{s}}{L_{AdS}}\right )^{1/3}\left (\frac{l_{p}^{2}}{m^{2}L_{AdS}^{5}}\right )\times\begin{cases}
+493& {\rm scalar} \\
-40& {\rm fermion} \\
-201& {\rm vector} 
\end{cases}\ ,\label{THLargeBH}
\end{align}
where the first term in \eqref{HorizonsLargeBH} and \eqref{THLargeBH} is only the leading part of GR answer and the second term is the leading correction from the EFT operators.

The corrected entropy is
\begin{align}
S&=8 \pi ^2  \left (\frac{L_{AdS}^{2}r_{s}}{l_{p}^{3}}\right )^{2/3}-\left (384a_{1}+96a_2\right)\pi^{2}\left (\frac{r_{s}}{L_{AdS}}\right )^{2/3} \nn
&\quad+\frac{1}{15120}\left (\frac{1}{mL_{AdS}}\right )^{2}\left (\frac{r_{s}}{L_{AdS}}\right )^{2/3}\times\begin{cases}
+1283& {\rm scalar} \\
-386& {\rm fermion} \\
-15& {\rm vector} 
\end{cases}\label{EntropyLargeBH}
\end{align}
where the first term in \eqref{EntropyLargeBH} is the leading part of the GR answer, the second term is the leading correction from the $a_{i}$ EFT operators and the term on the second line is the leading correction from the six-derivative, $c_{i}$ operators (the $b_i$'s don't contribute at leading order).

Finally, we evaluate the corrections to the $AdS$ free energy \eqref{CorrectedAdSFreeEnergy}.  Up to higher order corrections, we find
\begin{align}
E_{\rm AdS}&=M-\left (48a_{1}+12a_{2}\right )M\left ( \frac{1}{M_{p}^{2}L_{\rm AdS}^{2}}\right )\nn
&\quad+\frac{1}{20160\pi^{2}}M\left (\frac{1}{m^{2}M_{p}^{2}L_{\rm AdS}^{4}}\right )\times\begin{cases}
+296& {\rm scalar} \\
-71& {\rm fermion} \\
-36& {\rm vector} 
\end{cases}\ .\label{CorrectedAdSFreeEnergyExplicit}
\end{align}

\subsection{Asymptotically $dS$ Spacetimes\label{Sec:AsymptoticallydSSpacetimes}}

Finally, we consider asymptotically $dS$ spacetimes.  We focus on the changes regarding the cosmological horizon, as the corrections to the BH horizon are essentially the same as in the previous section, for small black holes.  

Up to higher order corrections, cosmological horizon now occurs at
\begin{align}
r_{h}&= L_{dS}+\frac{1}{120960\pi^{2}}\left (\frac{l_{p}^{2}}{m^{2}L_{dS}^{3}} \right ) \times\begin{cases}
-370& {\rm scalar} \\
+31 & {\rm fermion} \\
+150& {\rm vector} 
\end{cases}\label{CosHorizonSmallBH}
\end{align} and the corrected Hawking temperature is given by
\begin{align}
T_{H}&=\frac{1}{2 \pi  L_{dS}}+\frac{1}{241920\pi^{3}}\left (\frac{l_{p}^{2}}{m^{2}L_{dS}^{5}}\right ) \times\begin{cases}
+370& {\rm scalar} \\
-31 & {\rm fermion} \\
-150& {\rm vector} 
\end{cases}\label{THCosHor}
\end{align}
where the first term in \eqref{CosHorizonSmallBH} and \eqref{THCosHor} is only the leading part of GR answer (for $SdS$) and the second term is the leading correction from the EFT operators.  Note that the Hawking temperature is fully dictated by the change to the effective $L_{\rm dS}$ radius, $T_{H}=(2\pi r_{h})^{-1}$.

The corrected entropy is
\begin{align}
S&=8 \pi ^2\left (\frac{L_{dS}}{l_{p}}\right )^{2} +32\pi^{2}\left (12a_{1}+3a_{2}+2a_{\rm GB}\right )+\frac{1}{3780}\left (\frac{1}{mL_{dS}}\right )^{2}\times\begin{cases}
+370& {\rm scalar} \\
-31 & {\rm fermion} \\
-150& {\rm vector} 
\end{cases}\label{EntropyCosHor}
\end{align}
where the first term in \eqref{EntropyCosHor} is the leading part of the GR answer, the second term is the leading correction from the $a_{i}$ EFT operators and the third term is the leading correction from the six-derivative, $c_{i}$ operators (the $b_i$'s don't contribute at leading order).

Finally, for cosmological purposes, the result \eqref{CosHorizonSmallBH} is perhaps better recast a perturbative correction to the Hubble constant.  From \eqref{LambdaEff}, we find that the usual coefficient of the $\sim r^{2}$ component of the metric has changed from  $H^{2}\equiv L_{dS}^{-2}=\frac{\Lambda}{3M_{p}^{2}}$ to an effective value $H_{\rm eff}^{2}$ given by
\begin{align}
H_{\rm eff, \ scalars}^{2}&=\frac{\Lambda}{3M_{p}^{2}}+\frac{1}{544320\pi^{2}}\left (\frac{\Lambda}{3M_{p}^{2}}\right )\left (\frac{\Lambda^{2}}{m^{2}M_{p}^{6}}\right )\times\begin{cases}
+370& {\rm scalar} \\
-31& {\rm fermion} \\
-150& {\rm vector} 
\end{cases}\ .\label{HubbleEffective}
\end{align}
The corrections are interesting, but miniscule, as we discuss in Sec.$\!$ \ref{Sec:CorrectionsToHubbleDiscussion}.

\section{Interpretation\label{Sec:Interpretation}}

We now provide some discussion and interpretation regarding the results of Sec.$\!$ \ref{Sec:AllAsymptoticPossibilities}.  

\subsection{Corrections to Schwarzschild Horizon\label{Sec:CorrectionsToSchwHorizonDiscussion}}

There's a very natural interpretation for the signs which arise in the correction to the horizon distance of asymptotically flat black holes \eqref{HorizonsSmallBH}.  However, when probed in finer detail, the naive explanation appears to be much less convincing.  We present the naive interpretation and then discuss the ways in which it's lacking.

The temptation is to conclude that the sign on the correction is uniquely determined by the sign on the Planck Mass beta function \eqref{EFTBetaFunctions} which follows from vacuum polarization diagram for each particle species.    More conveniently, if we work with the beta function for Newton's constant, 
\begin{align}
\beta(G_{N})&=\frac{-\beta(M_{p}^{2})}{8\pi M_{p}^{4}}=\frac{m^{2}G_{N}^{2}}{6\pi}\times \begin{cases}
-1& {\rm scalar}\\
-2 &{\rm fermion}\\
+3& {\rm vector}
\end{cases}
\label{BetaGNewton}\ ,
\end{align}
by comparing against \eqref{HorizonsSmallBH} one finds that in theories where $\beta(G_{N})<0$, the horizon is slightly smaller than in pure GR, $r_{h}< r_{s}$, and vice versa for negative $\beta(G_{N})$.

 \begin{figure}[h!] 
  \captionsetup{width=0.9\textwidth}
   \centering
     \includegraphics[width=6in]{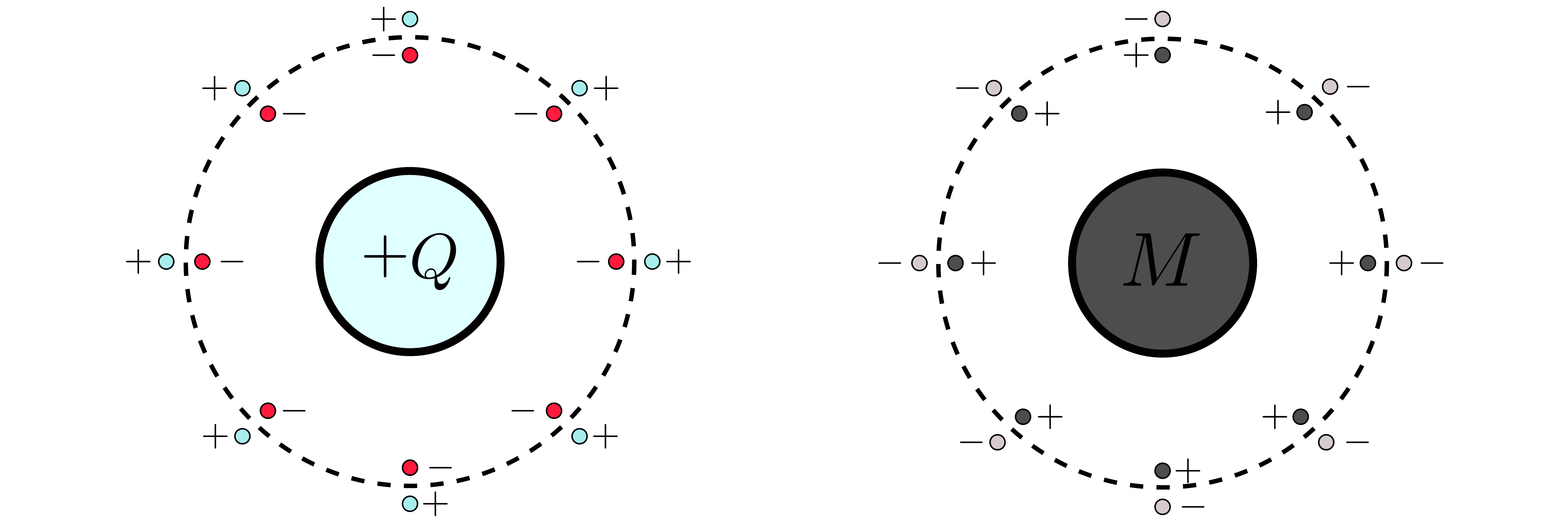}
   \caption{Left: Cartoon of electromagnetic vacuum polarization surround a charge $Q$. Right: Cartoon of gravitational vacuum polarization surrounding black hole of mass $M$.  Due to the virtual pairs surrounding the source, an observer sees an effective charge or mass determined not only by the large source, but also by the (net) number of virtual particles enclosed by their Gauss's law sphere.}
   \label{fig:GaussLawSpheres}
\end{figure}

This is exactly what one would naively expect.  The sign on $\beta(G_{N})$ determines whether gravity becomes weaker ($\beta(G_{N})<0$) or stronger ($\beta(G_{N})>0$) as we pass to shorter and shorter distance scales.  It is the direct analogue of the running of gauge couplings due to vacuum polarization by virtual particle-antiparticle pairs in gauge theories\footnote{The concept of virtual pairs polarized by gravitational fields is far more confusing  than its electromagnetic analogue.  See \cite{Dvali:2001gx} for a discussion of ``gravitational dipoles" consisting of positive and negative energy particles, inspiring the sketch in Fig.$\!$ \ref{fig:GaussLawSpheres}.}, see Fig.$\!$ \ref{fig:GaussLawSpheres}.  Very, very roughly an observer sitting a distance $r$ from the black hole experiences a potential $V\sim \frac{G_{N}(r)M}{r}$, where we've set $\mu=r^{-1}$.  Comparing to an equal mass black hole in pure GR, in which $G_{N}=\bar{G}_{N}={\rm constant}$, and tuning the EFT so that $G_{N}(r\to\infty)=\bar{G}_{N}$, an observer in an EFT with $\beta(G_{N})<0$ should feel a smaller potential at any given radius, as compared to GR: $G_{N}(r)M/r\le \bar{G}_{N}M/r$ with equality holding only at $r\to \infty$.  Thus, when $\beta(G_{N})<0$ gravitational fields are weaker, as compared to their GR counterparts, and the black hole horizon should occur at a smaller radius. Vice versa for $\beta(G_{N})>0$.

One should be extremely cautious about taking the above too seriously. When examined in greater detail, the explanation is left wanting in multiple ways:
\begin{itemize}
\item Heavy matter only contributes to the running of $G_{N}$ at energies $\mu \gtrsim m$ where the heavy field hasn't yet decoupled \cite{Manohar:1996cq}. We're instead probing physics at huge, macroscopic scales $\mu\lesssim  r_{s}^{-1}\ll\ll m$ where $G_{N}$ should be treated as constant and hence gleaning any long distance information from the sign of $\beta(G_{N})$ is an enormous extrapolation. 
\item The diagrams which determine the running and the corrections to the metric aren't the same: $\beta(G_{N})$ is determined solely through the standard vacuum polarization diagram with two external gravitons, whereas the $c_{i}$ EFT operators  \eqref{PreciseAction} responsible for the leading corrections to the Schwarzschild solution arise from triangle diagrams with three external gravitons.  
\item Studying the metric far from the black hole by expanding about flat space, $g_{\mu\nu}=\eta_{\mu\nu}+h_{\mu\nu}$, the leading long-distance corrections actually have the \textit{opposite} effect to what we previously inferred via the sign on $\beta(G_{N})$:
\begin{align}
h_{00}&=\frac{r_{s}}{r}+\frac{1}{3360\pi^{2}}\frac{r_{s}^{2}}{m^{2}M_{p}^{2}r^{6}}\times \begin{cases}+3 & {\rm scalar}\\
+2 &{\rm fermion}\\
-5 &{\rm vector}
\end{cases}+\ldots\ \label{LongDistanceh00CorrectedSchwarzschild}
\end{align}
Comparing to \eqref{BetaGNewton}, one finds that when $\beta(G_{N})<0$ the potential is actually \textit{larger} than its GR counterpart, and vice versa for $\beta(G_{N})>0$. 
 If the result \eqref{LongDistanceh00CorrectedSchwarzschild} persisted at all distances, we'd have found that negative (positive) $\beta(G_{N})$ corresponded to a larger (smaller) horizon, but the $\ldots$ in \eqref{LongDistanceh00CorrectedSchwarzschild} contains terms suppressed by factors of $r_{s}/r$ relative to the results shown. These are negligible far from the black hole, but crucial in the near horizon region $r\sim r_{s}$ and the numbers work out so that the result \eqref{HorizonsSmallBH} is obtained. 
\item Similar intuition leads us astray in the case of charged black holes in full QED.  The QED beta function is $\beta(e)=+\frac{e^{3}}{12\pi^{2}}>0$, meaning that one expects to feel a larger and larger effective charge as a source is approached.  Black holes with larger charges have smaller horizons and hence we expect the $\sim RFF$ and $\sim F^{4}$ operators generated by integrating out the electron to cause the black hole horizon to shrink, relative to its Einstein-Maxwell counterpart with the same asymptotic charge and mass.  Instead, the horizon grows \cite{Ruffini:2013hia,ToAppearWithKurt}.
\end{itemize} 

Therefore, the fact that the sign on the correction to the horizon radius tracks the sign on $\beta(M_{p})$ appears likely to be a coincidence rather than a deep, physical fact, despite the intuitive appeal of the naive argument.

\subsection{Temperature \label{Sec:TemperatureDiscussion}}

One expects the existence of heavy particles to affect Hawking temperatures through both on-shell and off-shell processes.  In the former case, additional species lead to additional channels that black holes can decay into.  This should raise the Hawking temperature.  In the latter case, virtual, off-shell effects alter the geometry directly, modifying the acceleration and redshift factors which determine the temperature felt by an observer at infinity, for the case of black hole horizons. The intuition for whether such off-shell effects should increase or decrease $T_{H}$ is far less clear. The mixture of signs on the corrections to $T_{H}$ (\eqref{THSmallBH}, \eqref{THLargeBH} and \eqref{THCosHor}) can be understood as a result of the fact that we've only calculated the off-shell effects (which are expected to dominate).

The signature of direct, on-shell particle creation is their non-perturbative, exponentially suppressed production.  The inherently perturbative calculations we've performed miss such effects completely.  The situation is analogous to the study of the Euler-Heisenberg action where the perturbative evaluation of the functional determinant or a matching calculation is sufficient for deriving the first few terms in the effective action,
\begin{align}
\mathcal{L}&=-\frac{1}{4e^{2}}F_{\mu\nu}^{2}+\frac{7}{90(4\pi)^{2}m^{4}}F^{\mu }{}_{\nu}F^{\nu}{}_{\rho}F^{\rho}{}_{\sigma}F^{\sigma}{}_{\mu }-\frac{1}{36(4\pi)^{2}m^{4}}\left (F_{\mu\nu}F^{\mu\nu}\right )^{2}+\ldots\ ,\label{EulerHeisenberg}
\end{align}
which describe the backreaction of virtual electrons on the photon field, but a non-perturbative calculation is required to calculate  the production rate of real  $e^{+}-e^{-}$ pairs in a constant electric field background $\Gamma\sim \exp\left (-m^{2}/eE\right )$ \cite{Schwinger:1951nm,Cohen:2008wz}. 

We've only calculated the analogue of \eqref{EulerHeisenberg} for our gravitational system, but this is expected to be enough.  The perturbative effects of virtual matter fields on the geometry are competing against direct production effects, but the production rate of such heavy particles should be suppressed by factors $\sim e^{-m/T_{H}}$ ($\ll 1$ by the EFT validity conditions \eqref{EFTValidityConditions}) and only a small fraction of these particles will be energetic enough to escape the black hole's potential well and get to infinity, where $T_{H}$ is felt.  In contrast, the suppression associated to virtual effects is only polynomial $\sim (mr_{s})^{-2}$, dominating the corrections in our regimes of interest.  As there is no clear expectation for whether such virtual effects should raise or lower Hawking temperatures, generally, it's perhaps not so surprising that there is little pattern to the signs on $\delta T_{H}$.

\subsection{Entropy\label{Sec:EntropyDiscussion}}

The most puzzling results are the corrections to the entropies.  We might have expected that horizons should carry more entropy in theories where there are more fields, simply because there is more data to keep track of. 
However, this expectation is not realized in the above findings.  Scalars increase the entropy of all horizons, fermions decrease the entropy of all horizons and vectors increase the entropy of flat black holes, but decrease the entropy for large $AdS$ black holes and $dS$ horizons; see \eqref{EntropySmallBH}, \eqref{EntropyCosHor} and \eqref{EntropyLargeBH}.

 It's possible that we are are not making the correct comparison between pure GR and the theories with heavy matter coupled in.  For instance, thinking in terms of entropy bounds \cite{Bekenstein:1980jp,Bousso:1999xy}, it is perhaps more sensible to compare the entropy of black holes at fixed \textit{size}, rather than at fixed mass.  In their simplest form, the bounds state that the entropy of a weakly gravitating system contained within sphere of some given surface area must be smaller than the entropy of a black hole with the same surface area (see \cite{Bousso:2002ju} for refinements and extensions of this criteria).  Higher derivative operators change the entropy of such black holes and we'd expect that the particular corrections which come from integrating out reasonable matter fields would allow us to store more information inside of a sphere of fixed size, simply from the increase in degrees of freedom, and so they should also increase the entropy of black holes at fixed horizon area. 
 
   However, explicit calculation shows that this is not always the case.  Happily one finds that scalars and fermions now both increase the entropy of flat black holes at fixed horizon area, but vectors now lead to a decrease.  From \eqref{BHHorizonShiftSmallRs}, we see that in order for a flat black hole in the EFT \eqref{PreciseAction} to have a horizon area $4\pi r_{s}^{2}$, its Schwarzschild radius must be $r'_{s}$ with
   \begin{align}
     r'_{s}=r_{s}+\left (10c_{1}+12c_{2}\right )\frac{l_{p}^{2}}{m^{2}r_{s}^{3}}+\ldots\ .
     \end{align}
     The corresponding entropy is
     \begin{align}
     S'&=8\pi \left (\frac{r_{s}}{l_{p}}\right )^{2}+192\pi^{2}\left (c_{1}+c_{2}\right )\frac{1}{m^{2}r_{s}^{2}}=8\pi \left (\frac{r_{s}}{l_{p}}\right )^{2}+\frac{1}{2520 m^{2}r_{s}^{2}}\times\begin{cases}
+19& {\rm scalar} \\
+8& {\rm fermion} \\
-27& {\rm vector} 
\end{cases}\label{FixedAreaEntropy}
     \end{align}
     where we evaluated $c_{1}$, $c_{2}$ on their values given in \eqref{EFTciValues} in the last line.   Note that the addition of non-minimal couplings such as $A_{\mu}A^{\mu}R$ or $A_{\mu}A_{\nu}R^{\mu\nu}$ would not change the result \eqref{FixedAreaEntropy} since they generate $\sim R^{3}$ terms built from the Ricci tensor and scalar alone, whereas it's the $c_{1}$, $c_{2}$ operators involving the full Riemann tensor which enter the above result.    This is reminiscent of results in \cite{Kabat:1995eq,Donnelly:2012st} where vectors also generated puzzling, negative contributions to entropy calculations (however, their findings are for massless vectors and fixed BH mass).
       If we simply \textit{demand} that the heavy matter fields increase the entropy of flat black holes at fixed horizon size, it imposes the following constraint on the massive spectrum
 \begin{align}
19 \sum_{\rm scalars}\frac{1}{m_{i}^{2}}+8\sum_{\rm fermions}\frac{1}{m_{i}^{2}}-27\sum_{\rm vectors}\frac{1}{m_{i}^{2}}>0\ , \label{PositiveEntropyBoundConstraint}
 \end{align}
 where only massive species enter the sum.  
 
   We could also revisit our renormalization conditions. We've chosen to tune parameters in both theories such that they share the same renormalized values of $M_{p}$ and $\Lambda$ at the low energies we are working at.  We could have instead demanded that the values of these parameters instead coincide in the deep UV at some energy scale $E_{\rm UV}\gg m$.  However, as discussed previously, if we imposed such a condition, the solutions in the two theories would no longer agree in the $m\to \infty$ limit and, in any case, this wouldn't appear to solve the problem.  In this scenario, the biggest correction to the Schwarzschild entropy comes from the running of the Planck mass (due to its role in the area law term).  Then, for instance, in the cases where a heavy scalar or fermion is minimally coupled to GR, we have $\beta(M_{p}^{2})>0$ and hence the IR value of $M_{p}^{2}$ is \textit{smaller} in the theory coupled to matter than it is in the pure GR theory, leading to a correspondingly smaller entropy than in pure GR once again.

\subsection{Corrections to Hubble\label{Sec:CorrectionsToHubbleDiscussion}}

Perhaps the most tantalizing correction to asymptotically $dS$ spacetimes is the change in the effective Hubble scale.  Higher derivative operators have changed $H^{2}\to H_{\rm eff}^{2}$ with
\begin{align}
H_{\rm eff}^{2}&=\frac{\Lambda}{3M_{p}^{2}}\times \left (1+c \frac{\Lambda^{2}}{m^{2}M_{p}^{6}}\right )\ ,\label{SchematicEffectiveHubble}
\end{align}
where $c$ is an $\mathcal{O}(1)$ number determined by the matter content \eqref{HubbleEffective}.  	Heavy fields slightly change the relation between the Hubble scale and the cosmological constant.

This effect is extremely tiny for particles such as the electron, for instance. Using $\Lambda\sim H^{2}M_{p}^{2}$ with $H\sim 70 \, {\rm km}\, {\rm s}^{-1}\, {\rm Mpc}^{-1}$, the fractional correction to Hubble is $\delta H/H\sim \mathcal{O}\left (\frac{H^{4}}{m^{2}M_{p}^{2}}\right )$ which is $\mathcal{O}\left (10^{-199}\right )$ for $M_{p}\sim 10^{19}{\rm GeV}$ and $m\sim .5 \, {\rm MeV}$.    In fact, as discussed in the next section, the correction \eqref{SchematicEffectiveHubble} may entirely be an artifact of the EFT treatment.

Still, it's quite tempting to speculate that such effects could be important in understanding the mismatch between the Hubble scale of our observed universe and its expected, natural size.  Though the small mass limit of \eqref{SchematicEffectiveHubble} is clearly unreliable and outside the validity of the EFT, there must exist \textit{some} expression which captures how fields lighter than $m \ll \Lambda/M_{p}^{3}$ affect $H_{\rm eff}$, asymptoting to \eqref{SchematicEffectiveHubble} in the appropriate regime. A better understanding of this limit is clearly desirable as it naively appears to indicate an $\mathcal{O}(1)$ change in the relationship between $\Lambda$ and the Hubble scale.

\section{Validity of Results: EFT Resolving Power\label{Sec:ValidityEFT}}

The corrections we've calculated are extremely small.  So small, in fact, that one worry whether they are large enough to constitute genuine predictions of the EFT. We find that indeed many of the results are so miniscule as to be untrustworthy.

 Every EFT comes with a resolution scale set by the cutoff, $\sim m$ in our case, and the result of an EFT calculation is only credible if it predicts phenomena which is resolvable given this scale.  Said a different way, the effective theory is built by averaging over or integrating out the physics at distance scales smaller than $\sim m^{-1}$, and hence the EFT should only be used to ask questions pertaining to length scales $\gg m^{-1}$.

This understanding is what allows us to cope with the higher derivatives in the first place.  Integrating out a field of mass $m$ generates a tower of $\sim R\left (\nabla^{2}/m^{2}\right )^{n}R$ operators which, if taken at face value, indicate the existence of catastrophic ghost instabilities \cite{Ostrogradsky:1850,Woodard:2006nt}, due to the resulting equations of motion with more than two time derivatives.  However, each operator is associated to a ghost whose mass is parametrically larger than the cutoff:
\begin{align}
 m_{\rm ghost}\sim m \left (M_{p}/m\right )^{1/(n+1)}\ge m\ .\label{GhostMasses}
 \end{align} 
  The result \eqref{GhostMasses} indicates the existence of problems only in regimes where the EFT is invalid anyway, distances $\sim m_{\rm ghost}^{-1}\lesssim m^{-1}$, so they are not of concern. Everything is under control as long as the higher derivative operators are treated perturbatively  \cite{Simon:1990ic,Jaen:1986iz,Burgess:2014lwa}.    Similar reasoning can be used to resolve other oddities which arise in effective theories.  For instance, it can be shown that the superluminal dispersion relations found in the low energy QED EFT around gravitational backgrounds \cite{Drummond:1979pp} are simple artifacts of EFT approximations as they cannot be used to create distance advances larger than the inverse cutoff \cite{Goon:2016une}.

The small shifts to horizon radii are artifacts of the above type.  Phrased as changes to the proper horizon area, we find
\begin{align}
\delta A_{\rm Schw.}\sim \frac{1}{m^{2}}\left (\frac{l_{p}^{2}}{r_{s}^{2}}\right )\ , \quad \delta A_{AdS}\sim \frac{1}{m^{2}}\left (\frac{l_{p}^{3}r_{s}}{L_{AdS}^{4}}\right )^{2/3}\ , \quad \delta A_{dS}\sim \frac{1}{m^{2}}\left (\frac{l_{p}^{2}}{L_{dS}^{2}}\right )\ ,
\end{align} 
parametrically, for the Schwarzschild BH, $dS$ cosmological horizon and large $AdS$ black hole, respectively.  It is natural to compare these to the minimal area built from the EFT cutoff: $A_{\rm min}\sim 1/m^{2}$.  In all cases  $\delta A/A_{\rm min}\ll 1$.  This is manifestly true for the Schwarzschild and $dS$ cases due to the hierarchy $L_{dS}, r_{s}\ll l_{p}$.   In the $AdS$ case it naively looks like we can create a macroscopic area increase $\delta A_{AdS}\gg m^{-2}$ by taking $r_{s}\gg L_{\rm AdS}^{4}/l_{p}^{3}$.  However, in this limit the $AdS$ Hawking temperature \eqref{THLargeBH} is super Planckian, $T_{H}\gg M_{p}$, so EFT calculations are not reliable.

The changes to the Hawking temperature are similarly suspect.  The typical wavelength of a Hawking quanta is $\lambda\sim T_{H}^{-1}$ and the shifts induced by heavy fields, $\delta \lambda\sim \delta T_{H}/\bar{T}_{H}^{2}$, are easily calculated,
\begin{align}
\delta \lambda_{\rm Schw.}\sim \frac{1}{m}\left (\frac{l_{p}^{2}}{mr_{s}^{3}}\right )\ , \quad \delta \lambda_{AdS}\sim \frac{1}{m}\left (\frac{l_{p}^{6}}{m^{3}L_{AdS}^{8}r_{s}}\right )^{1/3}\ , \quad \delta \lambda_{dS}\sim \frac{1}{m}\left (\frac{l_{p}^{2}}{mL_{dS}^{3}}\right )\ .
\end{align}
Each change is much smaller than the resolution scale, $\delta \lambda\ll m^{-1}$,
given the EFT conditions \eqref{EFTValidityConditions}.
Since the corrections to the entropy arise in part from corrections to the horizon radii and Hawking temperature, they may also be artifacts.  However, it's not entirely clear what the entropy shifts $\delta S$ should be compared to, as there's no natural length or energy scale associated to entropy.

It's interesting to note that similar conclusions would seem to apply to the effective field theory of GR \cite{Donoghue:1994dn} where no matter fields are included at all.  In the EFT of GR, the classical action is written as an expansion in powers of $R/M_{p}^{2}$ and $\nabla^{2}/M_{p}^{2}$, i.e. its cutoff is $\sim M_{p}$.  At one-loop, gravitons generate non-local operators $\sim R\ln(\square) R$ which dominate the corrections to the Schwarzschild metric at long distances $r\gg r_{s}$, generating a shift $\delta g_{\mu\nu}\sim \frac{r_{s}}{r}\left (\frac{l_{p}}{r}\right )^{2}$ \cite{BjerrumBohr:2002ks}.  Extrapolating to the near horizon regime, one expects loop corrections to shift the horizon radius by an amount $\delta r_{h}\sim l_{p}^{2}/r_{s}$ and hence the shifts to the horizon area and Hawking wavelength are $\delta A\sim l_{p}^{2}$ and $\delta\lambda \sim l_{p}^{2}/r_{s}$.  Neither shift falls squarely in the range of scales where the EFT is valid, so they should not be trusted, either, if the above estimates are correct. Of course, we are assuming here that the near horizon region is well-described by EFT, which may not be true \cite{Almheiri:2012rt}.  Note that other one-loop predictions such as corrections to the bending of light around massive bodies \cite{Bjerrum-Bohr:2016hpa} are expected to be resolvable, in contrast to the above.

Therefore, we find that many of the properties of horizons are remarkably resilient to higher derivative gravitational corrections. Estimates indicate that the above conclusions also hold in higher dimensions.    This appears to be a consequence of the weakness of gravity, rather than a fundamental property of horizons.   Local EFT corrections \textit{can} generate macroscopic shifts if additional forces are turned on.     As an example, one can introduce a $U(1)$ gauge field and study the QED EFT where the electron is integrated out. The higher derivative corrections in this theory can shift the area of the outer horizon of nearly extremally charged black holes by an amount parametrically larger than $1/m^{2}$ \cite{ToAppearWithKurt,Ruffini:2013hia}.  Alternatively, one can partially overcome the weakness of gravity by introducing an enormous number of species\footnote{We thank Joan Camps and Austin Joyce for pointing out this possibility.}.

\section{Conclusion\label{Sec:Conclusion}}

In this paper, we've found the perturbative corrections induced by loops of heavy matter fields on fundamental four-dimensional gravitational solutions: the Schwarzschild and Schwarzschild-$(A)dS$ spacetimes.  These are found by minimally coupling scalars, fermions and vectors to GR, integrating out the matter and studying solutions in the resulting higher derivative gravitational theory.  Such an EFT computation is valid when the Compton wavelength of the particle, $m^{-1}$, is much smaller than the typical curvature distance scale in the system under study.   In generic dimensions, $\sim R^{2}$ operators would dominate the corrections, but in $d=4$ they become subdominant to cubic operators $\sim R^{3}$ whose coefficients we calculate. 

After finding the perturbatively corrected solutions, we examined how the basic data about each space time was changed: the horizon radius, Hawking radiation and entropy.  For large $AdS$ black holes, we've also calculated the mass of the spacetime via the first law. The replica trick, as phrased by Lewkowycz and Maldacena \cite{Lewkowycz:2013nqa}, was used to calculate entropies (see Appendix \ref{Appendix:dSReplicaTrick} for an explicit demonstration of the method applied to $dS_{4}$).  For complicated higher derivative actions such as those we've studied, the replica trick proves to be a very efficient calculational technique.  We use it to find an expression for the entropy of static, spherically symmetric solutions to the most general, four-dimensional gravitational action containing terms involving up to six derivatives of the metric.

No general patterns were found amongst the corrections.  Depending on the type of matter and the spacetime, the defining properties of the solutions could either be increased or decreased, in accord with the findings of \cite{Matyjasek:2006nu} whose work we've extended. Whether heavy spin-$s$ fields increase or decrease various properties of the horizons of flat Schwarzschild black holes, large $AdS$ black holes and $dS$ cosmological horizons is indicated in Table \ref{Table:SignsOfCorrections} (see Sec.$\!$ \ref{Sec:AllAsymptoticPossibilities} for the complete expressions). While the above table is calculated at fixed $r_{s}$, it is also interesting to instead compare flat black holes at fixed horizon area, when considering entropy bounds. A particularly surprising finding in this setup is that massive vectors fields \textit{decrease} the entropy of such flat black holes (scalars and fermions lead to an increase), indicating that vector fields decrease the amount of information which can be contained in a sphere of fixed area.

\begin{table}
\begin{center}
\begin{tabular}{| c !{\vrule width0.8pt} c | c | c !{\vrule width0.8pt} c|}
\hline
 & Flat  & $AdS_{4}$ & $dS_{4}$  & $s$\\ \Xhline{2\arrayrulewidth}
 & $-$ & $-$ & $-$ & $0$\\ \cline{2-5}
$r_{h}$ & $-$ & $-$ & $+$  & $1/2$\\ \cline{2-5}
 & $+$ & $+$ & $+$  &$1$ \\  \Xhline{2\arrayrulewidth}
 & $+$ & $+$ & $+$ & $0$\\ \cline{2-5}
$T_{H}$ & $-$ & $-$ & $-$  &$1/2$ \\ \cline{2-5}
 & $+$& $-$ & $-$ & $1$\\  \Xhline{2\arrayrulewidth}
 & $+$ & $+$ & $+$ & $0$\\ \cline{2-5}
$S$ & $-$& $-$ & $-$  & $1/2$\\ \cline{2-5}
 & $+$ & $-$ & $-$  &$1$ \\ \hline
\end{tabular}
\caption{Collecting the signs of the corrections to the horizon radii, Hawking temperatures and entropies of asymptotically flat BH horizons, large $AdS_{4}$ BH horizons and $dS_{4}$ cosmological horizons arising from the influence of a heavy spin-$s$ particle. A ``$+$" or  ``$-$" indicates that the quantity is increased or decreased, respectively.   $M_{p}$, $\Lambda$ and $r_{s}$ are being held fixed in the above comparison.}
\label{Table:SignsOfCorrections}
\end{center}
\end{table}

However, many of these corrections should not be taken seriously as they are so miniscule\footnote{To get a sense for the scales of corrections in our universe, we calculate the electron-induced corrections to the $62M_{\odot}$ black hole which resulted from the LIGO GW150914 event in Appendix \ref{Appendix:LIGO}.}  as to be outside the resolving power of the effective field theory.  In all cases the shifts to the wavelength associated to Hawking radiation are far smaller than the heavy particle's Compton wavelength, i.e. the EFT cutoff $\sim m^{-1}$, and the changes in the horizon areas are much smaller than $1/m^{2}$.  Shifts to horizon entropies are lilliputian, too, but it is not as obvious whether these corrections are artifacts since it's unclear what to compare the corrections to.  The properties of spacetime are quite resilient against the effects of higher derivative gravitational corrections.

There are various avenues for future work.  Relatively simple extensions would involve including the effects of non-minimal matter couplings or moving to other dimensions.  A holographic interpretation of the asymptotically $AdS$ corrections could also be illuminating.  More generally, it would be interesting to study how constraints on generic quantum field theories translate into properties of corrected gravitational solutions.   For instance, unitarity and analyticity place conditions on the coefficients of various higher derivative gravitational operators\footnote{The results of \cite{Bellazzini:2015cra} are not immediately applicable, as they constrain the coefficients of $\sim R^{4}$ operators whereas $\sim R^{3}$ terms provide the dominant corrections to solutions, as we've seen, so an extension of \cite{Bellazzini:2015cra} would be required.} \cite{Bellazzini:2015cra} and it would be nice to understand these restrictions in terms of their effects on gravitational solutions.

A far more ambitious undertaking would be the computation of the black hole corrections which arise from loops of \textit{massless} particles, for example photons and the graviton itself.  Such loops lead to non-local operators which actually dominate over those we've considered in this paper.  For instance, the leading far-field metric correction induced by graviton loops is  $\delta g_{\mu\nu}\sim \frac{r_{s}}{r}\left (\frac{1}{M_{p}r}\right )^{2}$ \cite{Duff:1974ud,Donoghue:1994dn} and hence we expect massless loops to generate $\mathcal{O}\left ((M_{p}^{2}r_{s}^{2})^{-1} \right )$ corrections to the horizon of Schwarzschild black holes.  Meanwhile, heavy fields only correct the near horizon Schwarzschild metric by factors of $\mathcal{O}\left ((m^{2}M_{p}^{2}r_{s}^{4})^{-1}\right )\ll \mathcal{O}\left ((M_{p}^{2}r_{s}^{2})^{-1} \right )$ \eqref{PerturbedMetricExplicit2}.  However, despite their enhanced size, many of the graviton-loop induced corrections may still be untrustworthy in the EFT sense; see Sec.$\!$ \ref{Sec:ValidityEFT}.  A full understanding of such a calculation requires wrestling with subtle topics such as the non-linear treatment of non-local operators such as $\sim R\log\square R$ in curved space  \cite{El-Menoufi:2015cqw,Donoghue:2015nba} and is left to future work.

\noindent
{\bf Acknowledgments}:  We thank Daniel Baumann, Joan Camps, Siavash Golkar, Kurt Hinterbichler, Austin Joyce, Guilherme Pimentel and Mark Trodden for useful discussions and comments on the draft.  This work was supported by a Starting Grant of the European Research Council (ERC StG grant 279617) and the Delta ITP consortium, a program of the Netherlands Organization for Scientific Research (NWO), which is funded by the Dutch Ministry of Education, Culture and Science (OCW).

\appendix

\section{\label{Appendix:dSReplicaTrick} de Sitter Entropy via the Replica Trick}

In this Appendix, we explicitly calculate the four-dimensional $dS$ entropy using the replica trick.

The Euclidean lagrangian is
\begin{align}
\mathcal{L}_{E}[g]\equiv -\frac{M_{p}^{2}}{2}\left (R-\frac{6}{L_{dS}^{2}}\right )\label{EuclideanLagrangianAppendix}
\end{align} and the usual Euclidean action and smooth metric solution are given by
\begin{align}
I_{E}^{(1)}&=\int_{0}^{2\pi L_{dS}}\rd \tau\int\rd^{3}x\sqrt{\bar{g}^{(1)}}\, \mathcal{L}_{E}[\bar g^{(1)}] \nn
\bar{g}_{\mu\nu}^{(1)}\rd x^{\mu}\rd x^{\nu}&=\left (1-r^{2}/L_{dS}^{2}\right )\rd \tau^{2}+\left (1-r^{2}/L_{dS}^{2}\right )^{-1}\rd r^{2}+r^{2}\rd \Omega^{2}\ ,\label{EuclideandSAppendix}
\end{align}
where we identity $\tau\sim \tau+2\pi L_{dS}$ and $\int \rd^{3}x$ is the integral over $\{r,\theta,\phi\}$ with $0\le r\le L_{dS}$.  The replica action, with $n=1+\epsilon$, is similarly given by
\begin{align}
I_{E}^{(1+\epsilon)}&=\int_{0}^{(1+\epsilon)2\pi L_{dS}}\rd \tau\int\rd^{3}x\sqrt{\bar	g^{(1+\epsilon)}}\, \mathcal{L}_{E}[\bar	g^{(1+\epsilon)}]\nn
&=\int_{0}^{(1+\epsilon)2\pi L_{dS}}\rd \tau\int\rd^{3}x\sqrt{g^{\rm off}}\, \mathcal{L}_{E}[g^{\rm off}]+\mathcal{O}(\epsilon^{2})
\end{align}
and we take the off-shell metric to be of the general form \eqref{GeneralOffShellEuclideanizedGenericSphericalMetric}
\begin{align}
g^{\rm off}_{\mu\nu}\rd x^{\mu}\rd x^{\nu}&=\frac{\left (1-r^{2}/L_{dS}^{2}\right )}{(1+\varepsilon(r))^{2-\alpha}}\rd \tau^{2}+\frac{(1+\varepsilon(r))^{\alpha}}{\left (1-r^{2}/L_{dS}^{2}\right )}\rd r^{2}+r^{2}\rd \Omega^{2}\ ,\label{ReplicadSSolnAppendix}
\end{align}
where we identify $\tau\sim \tau+(1+\epsilon)2\pi L_{dS}$ in the replica spacetime \eqref{ReplicadSSolnAppendix}.

It is straightforward to evaluate the two integrands via computer:
\begin{align}
\sqrt{\bar{g}^{(1)}}\mathcal{L}_{E}\left [\bar{g}^{(1)}\right ]&=-r^{2}\sin\theta \frac{3M_{p}^{2}}{L_{dS}^{2}}\nn
\sqrt{g^{\rm off}}\mathcal{L}_{E}\left [g^{\rm off}\right ]&=-r^{2}\sin\theta \frac{3M_{p}^{2}}{L_{dS}^{2}}\Big[1+   \varepsilon \left (\frac{\alpha L_{dS}^{2}-3(1+\alpha)r^{2}}{3r^{2}}\right )+ \varepsilon' \left (\frac{ 2 L_{dS}^2+(\alpha -5)
   r^2}{3 r}\right )\nn
   &\quad +   \varepsilon''\left (\frac{(2-\alpha)\left(L_{dS}^2-r^2\right) }{6}   \right )
\Big]+\mathcal{O}(\epsilon^{2})\ ,
\end{align}
where primes indicate radial derivatives and we've left the argument of $\varepsilon(r)$ implicit.

As $dS$ is compact, there are no boundary terms in the action and the only restriction on $\varepsilon(r)$ is that it not introduce singularities anywhere in the spacetime.    In particular, we require $\varepsilon(L_{dS})=\epsilon$ and $\varepsilon(0)=0$ to avoid deficit angles at the cosmological horizon and the origin, respectively, and we also assume that there are no singularities in $\varepsilon$ or its derivatives.

  Writing $\varepsilon(r)=\epsilon \psi(r)$ where $\psi(r)$ obeys $\psi(L_{dS})=1$, $\psi(0)=0$, we can use the above results to reduce the $dS$ entropy, calculated via \eqref{EntropyInTermsOfActionsAndEpsilon}, to a radial integral
\begin{align}
S_{dS}&=8\pi^{2}M_{p}^{2}L_{dS}^{2}\int_{0}^{L_{dS}}\rd r\,\left [\psi\frac{3 r^2  }{L_{dS}^3}+\psi '\left(\frac{5 r^3}{L_{dS}^3}-\frac{2 r}{L_{dS}}\right) + \psi ''\left(\frac{r^4}{L_{dS}^3}-\frac{r^2}{L_{dS}}\right) \right ]\nn
&\quad +8\pi^{2}M_{p}^{2}L_{dS}^{2} \alpha\int_{0}^{L_{dS}}\rd r\, \left [\psi\left(\frac{3 r^2}{L_{dS}^3}-\frac{1}{L_{dS}}\right)  - \psi '\frac{r^3}{L_{dS}^3}+\psi
   ''\left(\frac{r^2}{2 L_{dS}}- \frac{r^4}{2 L_{dS}^3}\right) \right ]\label{dSEntropyAsRadialIntegral1Appendix}\ .
\end{align}
  Since the entropy shouldn't depend on the precise details of the deforming function $\psi$, we should be able to write \eqref{dSEntropyAsRadialIntegral1Appendix} as an integral of total derivatives and, indeed, we can
\begin{align}
S_{dS}
&=8\pi^{2}M_{p}^{2}L_{dS}^{2}\int_{0}^{L_{dS}}\rd r\,\partial_{r}\left (-\frac{r^{2}\psi'}{L_{dS}}+\frac{r^{3}\partial_{r}(r\psi)}{L_{dS}^{3}} +\frac{\alpha r^{4}\partial_{r}\left (r^{-2}\psi\right )}{2L_{dS}}-\frac{\alpha r^{6}\partial_{r}\left (r^{-2}\psi\right )}{2L_{dS}^{3}}\right )\nn
&=8\pi^{2}M_{p}^{2}L_{dS}^{2}\left [-\frac{r^{2}\psi'}{L_{dS}}+\frac{r^{4}\psi'+r^{3}\psi}{L_{dS}^{3}}+\alpha\left (\frac{r^{2}\psi'-2r\psi}{2L_{dS}}-\frac{r^{4}\psi'-2r^{3}\psi}{2L_{dS}^{3}}\right )\right ]\Big|_{0}^{L_{dS}}\ .
\end{align}
Evaluating for any $\psi(r)$ obeying the previously stated conditions, the $\alpha$-dependent terms cancel out and we find the usual area law
\begin{align}
S_{dS}&= 8\pi^{2}M_{p}^{2}L_{dS}^{2}=\frac{A}{4G_{N}}\ .
\end{align}
The generalization to arbitrary dimensions should be straightforward.

\section{General Entropy for $g_{\tau\tau}\neq 1/ g_{rr}$\label{Appendix:GeneralEntropyFormula}}

Below is the unilluminating entropy formula for a generic spherically symmetric metric of the form 
\begin{align}
\bar{g}^{(1)}_{\mu\nu}\rd x^{\mu}\rd x^{\nu}&=f(r)\rd \tau^{2}+\frac{1}{g(r)}\rd r^{2}+r^{2}\rd \Omega^{2}\ ,
\end{align} given the action \eqref{PreciseAction}:
\begin{align}
S&= 8 \pi ^2\left (\frac{r_{h}}{l_{p}}\right )^{2}  \nn
&\quad +a_{1}\Big[ -\frac{24 \pi ^2 r_h ^2 f'' g'}{f'}-8 \pi ^2 r_h ^2 g''-128 \pi ^2 r_h  g'+64 \pi ^2  \Big]\nn
&\quad +a_{2}\Big[ -\frac{12 \pi ^2 r_h ^2 f'' g'}{f'}-4 \pi ^2 r_h ^2 g''-32 \pi ^2 r_h  g'  \Big]\nn
&\quad +64 \pi ^2 a_{\rm GB}   \nn
&\quad +\frac{b_{1}}{m^{2}}\Big[   -\frac{80 \pi ^2 r_h ^2 f^{(3)} g'^2}{3 f'}+\frac{16 \pi ^2 r_h ^2 f''^2 g'^2}{f'^2}-\frac{32 \pi ^2
   r_h  f'' g'^2}{f'}-\frac{16 \pi ^2 r_h ^2 f'' g' g''}{f'}+64 \pi ^2
   g'^2\nn
   &\quad-\frac{128 \pi ^2 g'}{r_h }-\frac{16}{3} \pi ^2 r_h ^2 g^{(3)} g'-96 \pi ^2 r_h  g' g'' \Big]\nn
&\quad +\frac{b_{2}}{m^{2}}\Big[   -\frac{40 \pi ^2 r_h ^2 f^{(3)} g'^2}{3 f'}+\frac{8 \pi ^2 r_h ^2 f''^2 g'^2}{f'^2}-\frac{8 \pi ^2
   r_h  f'' g'^2}{f'}-\frac{8 \pi ^2 r_h ^2 f'' g' g''}{f'}+32 \pi ^2
   g'^2\nn
   &\quad -\frac{8}{3} \pi ^2 r_h ^2 g^{(3)} g'-24 \pi ^2 r_h  g' g'' \Big]\nn
&\quad +\frac{c_{1}}{m^{2}}\Big[  \frac{27 \pi ^2 r_h ^2 f''^2 g'^2}{f'^2}+\frac{18 \pi ^2 r_h ^2 f'' g' g''}{f'}+3 \pi ^2
   r_h ^2 g''^2  \Big]\nn
&\quad +\frac{c_{2}}{m^{2}}\Big[   \frac{27 \pi ^2 r_h ^2 f''^2 g'^2}{f'^2}+\frac{96 \pi ^2 r_h  f'' g'^2}{f'}-\frac{48 \pi ^2 f''
   g'}{f'}+\frac{18 \pi ^2 r_h ^2 f'' g' g''}{f'}+3 \pi ^2 r_h ^2 g''^2-16 \pi ^2
   g''\nn
   &\quad +64 \pi ^2 g'^2+32 \pi ^2 r_h  g' g''+\frac{64 \pi ^2}{r_h ^2} \Big]\nn
&\quad +\frac{c_{3}}{m^{2}}\Big[  \frac{27 \pi ^2 r_h ^2 f''^2 g'^2}{f'^2}+\frac{288 \pi ^2 r_h  f'' g'^2}{f'}-\frac{144 \pi ^2
   f'' g'}{f'}+\frac{18 \pi ^2 r_h ^2 f'' g' g''}{f'}+3 \pi ^2 r_h ^2 g''^2-48 \pi
   ^2 g''\nn
   &\quad +768 \pi ^2 g'^2-\frac{768 \pi ^2 g'}{r_h }+96 \pi ^2 r_h  g' g''+\frac{192 \pi ^2}{r_h ^2}  \Big]\nn
&\quad +\frac{c_{4}}{m^{2}}\Big[ \frac{27 \pi ^2 r_h ^2 f''^2 g'^2}{4 f'^2}+\frac{36 \pi ^2 r_h  f'' g'^2}{f'}+\frac{9 \pi ^2 r_h ^2
   f'' g' g''}{2 f'}+\frac{3}{4} \pi ^2 r_h ^2 g''^2+48 \pi ^2 g'^2+12 \pi ^2 r_h  g'
   g''   \Big]\nn
&\quad +\frac{c_{5}}{m^{2}}\Big[  \frac{27 \pi ^2 r_h ^2 f''^2 g'^2}{2 f'^2}+\frac{96 \pi ^2 r_h  f'' g'^2}{f'}-\frac{24 \pi ^2
   f'' g'}{f'}+\frac{9 \pi ^2 r_h ^2 f'' g' g''}{f'}+\frac{3}{2} \pi ^2 r_h ^2
   g''^2-8 \pi ^2 g''\nn
   &\quad +192 \pi ^2 g'^2-\frac{128 \pi ^2 g'}{r_h }+32 \pi ^2 r_h  g' g''+\frac{32 \pi
   ^2}{r_h ^2}  \Big]\nn
&\quad +\frac{c_{6}}{m^{2}}\Big[ \frac{27 \pi ^2 r_h ^2 f''^2 g'^2}{4 f'^2}+\frac{24 \pi ^2 r_h  f'' g'^2}{f'}+\frac{9 \pi ^2 r_h ^2
   f'' g' g''}{2 f'}+\frac{3}{4} \pi ^2 r_h ^2 g''^2+48 \pi ^2 g'^2\nn
   &\quad -\frac{32 \pi ^2 g'}{r_h }+8
   \pi ^2 r_h  g' g''   \Big]\ ,
\end{align}
included for completeness.

\section{Rules for Matrix Elements \label{Appendix:RulesForMatrixElements}}

In this Appendix, we collect the rules calculating functional determinants perturbatively.  This is the same method used in \cite{Goon:2016ihr}, though the notation is different; see that reference for more explicit calculations.  Here work in arbitrary $d$-dimensions and use conventions $\delta(x-y)=\delta^{d}(x-y)$, $\tilde{p}^{\mu}=p^{\mu}/2\pi$, $\tilde{\delta}(k-p)=(2\pi)^{d}\delta^{d}(k-p)$, $\langle x|p\rangle=e^{ip\cdot x}$ and $f(p)=\int\rd^{d}x\, e^{-ix\cdot p}f(x)$ so that $\mathbf{1}=\int\rd^{d}\tilde{p}|p\rangle\langle p|=\int\rd^{d}x\, |x\rangle\langle x|$.

Consider, for illustrative purposes, calculating a flat space functional determinant of the form
\begin{align}
S_{\rm eff}=-\Tr\ln \left (\partial^{2}-m^{2}-\mathcal{O}\right )\label{ExampleTrLnAppendix}
\end{align}
where $\mathcal{O}=\mathcal{O}(\phi,\partial\phi,\partial\ldots)$ is some operator built from, say, a scalar field $\phi$, derivatives thereof and also derivatives acting on the $\delta^{d}(x-y)$ which usually left implicit in the functional determinant.  For instance, if we couple a heavy, complex scalar $\Phi$ to $\phi$ via the action
\begin{align}
 S[\Phi,\phi]&=\int\rd^{d}x\, \left [-(1+\phi)|\partial\Phi|^{2}-\left (m^{2}+\phi^{2}+\frac{\partial^{2}\phi}{\Lambda}\right )|\Phi|^{2}\right ]\ ,
 \end{align} then the effective action we'd get from integrating out $\Phi$ is of the form \eqref{ExampleTrLnAppendix}:
 \begin{align}
  S_{\rm eff}&=-\Tr\ln \frac{\delta S[\Phi,\phi]}{\delta \Phi(x)\delta\Phi^{*}(y)}=-\Tr\ln\left (\partial^{2}-m^{2}+\phi^{2}+\frac{\partial^{2}\phi}{\Lambda}+\phi\partial^{2}+\partial_{\mu}\phi\partial^{\mu}\right )\delta^{d}(x-y)\ .
  \end{align} The functional determinant can then be calculated perturbatively by writing
\begin{align}
S_{\rm eff}&=-\Tr\ln\left (\mathbf{1}-\frac{1}{\partial^{2}-m^{2}}\mathcal{O}\right )\ ,
\end{align}
which is equivalent to \eqref{ExampleTrLnAppendix} up to a divergent, field-independent term, and expanding out the logarithm and performing the traces.
\begin{align}
S_{\rm eff}&=\Tr \left (\frac{1}{\partial^{2}-m^{2}}\mathcal{O}\right )+\frac{1}{2}\Tr \left (\frac{1}{\partial^{2}-m^{2}}\mathcal{O}\right )^{2}+\frac{1}{3}\Tr \left (\frac{1}{\partial^{2}-m^{2}}\mathcal{O}\right )^{3}+\ldots
\end{align}

  In taking the trace, one encounters a few basic types of matrix elements:
  \begin{itemize}
  \item The simplest case is when $\mathcal{O}$ is a polynomial in $\phi$.  The matrix elements of $\phi$ itself are:
  \begin{align}
  \langle x|\phi|y\rangle=\phi(x)\delta(x-y)\ , \quad \langle x|\phi|p\rangle=\phi(x)e^{ip\cdot x}\ , \quad \langle k|\phi|p\rangle=\phi(k-p)\ ,\label{SimplestMatrixElementAppendix}
  \end{align}
  and the matrix element of $\mathcal{O}(\phi)$ can be found by repeated insertions of unity.  
  \item Matrix elements involving derivatives acting to the right on the implicit delta function are only slightly more complicated.  The rules are:
  \begin{align}
\langle x|\partial_{\mu}|y\rangle=\partial_{\mu}^{(x)}\delta(x-y)\ , \quad \langle x|\partial_{\mu}|p\rangle=ip_{\mu}  e^{ip\cdot x} \ , \quad \langle k|\partial_{\mu}|p\rangle=  ip_{\mu}\tilde{\delta}(k-p) \ .\label{DerivativeOnImplicitDeltaMatrixElementAppendix}
\end{align}
where the superscript on $\partial^{(x)}_{\mu}$ indicates that the derivative is taken with respect to the $x$ argument.  In particular, the propagator comes from $\langle k|(\partial^{2}-m^{2})^{-1}|p\rangle=(-k^{2}-m^{2})^{-1}\tilde{\delta}(k-p)$.
\item Finally, matrix elements of derivatives acting on fields are somewhat tricky.  One needs to treat objects such as $\partial_{\mu}\phi$ as a single entity, yielding
\begin{align}
   \langle x|\partial_{\mu}\phi|y\rangle&\! =\! \partial_{\mu}\phi(x)\delta(x-y)\ , \ \langle x|\partial_{\mu}\phi|p\rangle\! =\! e^{ip\cdot x}\partial_{\mu}\phi(x)\ , \ \langle k|\partial_{\mu}\phi|p\rangle\! =\!  i(k-p)_{\mu}\phi(k-p)\ .\label{DerivativesOnAFieldMatrixElementAppendix}
   \end{align} 
   The potentially confusing subtlety is that one cannot insert factors of unity between the derivative and the field in expressions like $\partial_{\mu}\phi$, otherwise one is lead to inconsistent results.
\end{itemize}   In each of the above cases, matrix elements of the form $\langle x|\mathcal{O}|p\rangle$ and $\langle k|\mathcal{O}|p\rangle$ follow from the initial position space matrix element $\langle x|\mathcal{O}|y\rangle$ via completeness relations.

Any other desired matrix elements can be derived by using further completeness relations on the above expressions or from simple generalizations of \eqref{SimplestMatrixElementAppendix}, \eqref{DerivativeOnImplicitDeltaMatrixElementAppendix} and \eqref{DerivativesOnAFieldMatrixElementAppendix}.

\section{Electron Corrections to LIGO GW150914 Black Hole\label{Appendix:LIGO}}

In order to get a sense for the scales at hand, we can consider computing the electron-induced corrections to the end state black hole of the observed LIGO GW150914 merger \cite{Abbott:2016blz}.

The mass of the final black hole was determined to be $M\approx 62M_\odot\approx 6.9\times 10^{58}{\rm GeV}$.  Assuming pure GR and negligible spin, the black hole horizon, Hawking temperature and entropy are calculated to be 
\begin{align}
\bar r_{h}=1.8\times 10^{5}{\rm m}\ , \quad \bar{T}_{H}&= 8.5\times 10^{-23}{\rm GeV} \ , \quad \bar{S}=4.0\times 10^{80}\ ,
\end{align}
from the leading terms in \eqref{HorizonsSmallBH}, \eqref{THSmallBH} and \eqref{EntropySmallBH}.  Calculating the corrections in these expressions due to the electron, mass $m=5.1\times 10^{-4}{\rm GeV}$, we find
\begin{align}
 \delta r_{h}=-3.4\times 10^{-114}{\rm m}\ , \quad \delta T_{H}&= -2.5\times 10^{-142}{\rm GeV} \ , \quad \delta S=-1.18\times 10^{-39}\ .
\end{align}
As advertised, the relative corrections are puny. However, the change in the entropy can also be phrased as a change to the number of black hole microstates $N=\exp(S)$, in which case $\delta N$ is numerically enormous.

\bibliographystyle{utphys}
\bibliography{HeavyFieldsAndBlackHoles}

\providecommand{\href}[2]{#2}\begingroup\raggedright\begin{thebibliography}{10}

\bibitem{Ruffini:2013hia}
R.~Ruffini, Y.-B. Wu, and S.-S. Xue, ``{Einstein-Euler-Heisenberg Theory and
  charged black holes},''
  \href{http://dx.doi.org/10.1103/PhysRevD.88.085004}{{\em Phys. Rev.} {\bf
  D88} (2013)  085004},
\href{http://arxiv.org/abs/1307.4951}{{\tt arXiv:1307.4951 [hep-th]}}.

\bibitem{Lu:1993sq}
M.~Lu and M.~B. Wise, ``{Black holes with a generalized gravitational
  action},'' \href{http://dx.doi.org/10.1103/PhysRevD.47.R3095}{{\em Phys.
  Rev.} {\bf D47} (1993)  3095--3098},
\href{http://arxiv.org/abs/gr-qc/9301021}{{\tt arXiv:gr-qc/9301021 [gr-qc]}}.

\bibitem{Matyjasek:2006fq}
J.~Matyjasek, M.~Telecka, and D.~Tryniecki, ``{Higher dimensional black holes
  with a generalized gravitational action},''
  \href{http://dx.doi.org/10.1103/PhysRevD.73.124016}{{\em Phys. Rev.} {\bf
  D73} (2006)  124016},
\href{http://arxiv.org/abs/hep-th/0606254}{{\tt arXiv:hep-th/0606254
  [hep-th]}}.

\bibitem{Taylor:1999ic}
B.~E. Taylor, W.~A. Hiscock, and P.~R. Anderson, ``{Semiclassical charged black
  holes with a quantized massive scalar field},''
  \href{http://dx.doi.org/10.1103/PhysRevD.61.084021}{{\em Phys. Rev.} {\bf
  D61} (2000)  084021},
\href{http://arxiv.org/abs/gr-qc/9911119}{{\tt arXiv:gr-qc/9911119 [gr-qc]}}.

\bibitem{Kats:2006xp}
Y.~Kats, L.~Motl, and M.~Padi, ``{Higher-order corrections to mass-charge
  relation of extremal black holes},''
  \href{http://dx.doi.org/10.1088/1126-6708/2007/12/068}{{\em JHEP} {\bf 0712}
  (2007)  068},
\href{http://arxiv.org/abs/hep-th/0606100}{{\tt arXiv:hep-th/0606100
  [hep-th]}}.

\bibitem{Smolic:2013gz}
J.~Smolic and M.~Taylor, ``{Higher derivative effects for 4d AdS gravity},''
  \href{http://dx.doi.org/10.1007/JHEP06(2013)096}{{\em JHEP} {\bf 06} (2013)
  096},
\href{http://arxiv.org/abs/1301.5205}{{\tt arXiv:1301.5205 [hep-th]}}.

\bibitem{Matyjasek:2006nu}
J.~Matyjasek, ``{Entropy of quantum-corrected black holes},''
  \href{http://dx.doi.org/10.1103/PhysRevD.74.104030}{{\em Phys. Rev.} {\bf
  D74} (2006)  104030},
\href{http://arxiv.org/abs/gr-qc/0610020}{{\tt arXiv:gr-qc/0610020 [gr-qc]}}.

\bibitem{Carroll:2004st}
S.~M. Carroll, {\em {Spacetime and geometry: An introduction to general
  relativity}}.
\newblock 2004.
\newblock
\url{http://www.slac.stanford.edu/spires/find/books/www?cl=QC6:C37:2004}.
\newblock

\bibitem{Misner:1974qy}
C.~W. Misner, K.~S. Thorne, and J.~A. Wheeler, {\em {Gravitation}}.
\newblock W. H. Freeman, San Francisco,
1973.
\newblock

\bibitem{Edgar:2001vv}
S.~B. Edgar and A.~Hoglund, ``{Dimensionally dependent tensor identities by
  double antisymmetrization},'' \href{http://dx.doi.org/10.1063/1.1425428}{{\em
  J. Math. Phys.} {\bf 43} (2002)  659--677},
\href{http://arxiv.org/abs/gr-qc/0105066}{{\tt arXiv:gr-qc/0105066 [gr-qc]}}.

\bibitem{Fulling:1992vm}
S.~A. Fulling, R.~C. King, B.~G. Wybourne, and C.~J. Cummins, ``{Normal forms
  for tensor polynomials. 1: The Riemann tensor},''
\href{http://dx.doi.org/10.1088/0264-9381/9/5/003}{{\em Class. Quant. Grav.}
  {\bf 9} (1992)  1151--1197}.

\bibitem{'tHooft:1974bx}
G.~'t~Hooft and M.~J.~G. Veltman, ``{One loop divergencies in the theory of
  gravitation},''
{\em Annales Poincare Phys. Theor.} {\bf A20} (1974)  69--94.

\bibitem{Liko:2007vi}
T.~Liko, ``{Topological deformation of isolated horizons},''
  \href{http://dx.doi.org/10.1103/PhysRevD.77.064004}{{\em Phys. Rev.} {\bf
  D77} (2008)  064004},
\href{http://arxiv.org/abs/0705.1518}{{\tt arXiv:0705.1518 [gr-qc]}}.

\bibitem{Chatterjee:2013daa}
S.~Chatterjee and M.~Parikh, ``{The second law in four-dimensional
  Einstein-Gauss-Bonnet gravity},''
  \href{http://dx.doi.org/10.1088/0264-9381/31/15/155007}{{\em Class. Quant.
  Grav.} {\bf 31} (2014)  155007},
\href{http://arxiv.org/abs/1312.1323}{{\tt arXiv:1312.1323 [hep-th]}}.

\bibitem{Sarkar:2010xp}
S.~Sarkar and A.~C. Wall, ``{Second Law Violations in Lovelock Gravity for
  Black Hole Mergers},''
  \href{http://dx.doi.org/10.1103/PhysRevD.83.124048}{{\em Phys. Rev.} {\bf
  D83} (2011)  124048},
\href{http://arxiv.org/abs/1011.4988}{{\tt arXiv:1011.4988 [gr-qc]}}.

\bibitem{xAct}
J.~M. Mart\'in-Garc\'ia, {\em {xAct, Efficient tensor computer algebra for
  mathematica}}.
\newblock \url{http://www.xact.es}.

\bibitem{ToAppearWithKurt}
G.~Goon and K.~Hinterbichler, ``{In Preparation},''.

\bibitem{Ginsparg:1982rs}
P.~H. Ginsparg and M.~J. Perry, ``{Semiclassical Perdurance of de Sitter
  Space},''
\href{http://dx.doi.org/10.1016/0550-3213(83)90636-3}{{\em Nucl. Phys.} {\bf
  B222} (1983)  245--268}.

\bibitem{Bousso:1997wi}
R.~Bousso and S.~W. Hawking, ``{(Anti)evaporation of Schwarzschild-de Sitter
  black holes},'' \href{http://dx.doi.org/10.1103/PhysRevD.57.2436}{{\em Phys.
  Rev.} {\bf D57} (1998)  2436--2442},
\href{http://arxiv.org/abs/hep-th/9709224}{{\tt arXiv:hep-th/9709224
  [hep-th]}}.

\bibitem{Nojiri:1998ph}
S.~Nojiri and S.~D. Odintsov, ``{Quantum evolution of Schwarzschild-de Sitter
  (Nariai) black holes},''
  \href{http://dx.doi.org/10.1103/PhysRevD.59.044026}{{\em Phys. Rev.} {\bf
  D59} (1999)  044026},
\href{http://arxiv.org/abs/hep-th/9804033}{{\tt arXiv:hep-th/9804033
  [hep-th]}}.

\bibitem{Bytsenko:1998md}
A.~A. Bytsenko, S.~Nojiri, and S.~D. Odintsov, ``{Quantum generation of
  Schwarzschild-de Sitter (Nariai) black holes in effective Dilaton - Maxwell
  gravity},'' \href{http://dx.doi.org/10.1016/S0370-2693(98)01330-6}{{\em Phys.
  Lett.} {\bf B443} (1998)  121--126},
\href{http://arxiv.org/abs/hep-th/9808109}{{\tt arXiv:hep-th/9808109
  [hep-th]}}.

\bibitem{Nojiri:2013su}
S.~Nojiri and S.~D. Odintsov, ``{Anti-Evaporation of Schwarzschild-de Sitter
  Black Holes in $F(R)$ gravity},''
  \href{http://dx.doi.org/10.1088/0264-9381/30/12/125003}{{\em Class. Quant.
  Grav.} {\bf 30} (2013)  125003},
\href{http://arxiv.org/abs/1301.2775}{{\tt arXiv:1301.2775 [hep-th]}}.

\bibitem{Gibbons:1976ue}
G.~W. Gibbons and S.~W. Hawking, ``{Action Integrals and Partition Functions in
  Quantum Gravity},''
\href{http://dx.doi.org/10.1103/PhysRevD.15.2752}{{\em Phys. Rev.} {\bf D15}
  (1977)  2752--2756}.

\bibitem{Wald:1993nt}
R.~M. Wald, ``{Black hole entropy is the Noether charge},''
  \href{http://dx.doi.org/10.1103/PhysRevD.48.R3427}{{\em Phys. Rev.} {\bf D48}
  (1993) no.~8, R3427--R3431},
\href{http://arxiv.org/abs/gr-qc/9307038}{{\tt arXiv:gr-qc/9307038 [gr-qc]}}.

\bibitem{Iyer:1994ys}
V.~Iyer and R.~M. Wald, ``{Some properties of Noether charge and a proposal for
  dynamical black hole entropy},''
  \href{http://dx.doi.org/10.1103/PhysRevD.50.846}{{\em Phys. Rev.} {\bf D50}
  (1994)  846--864},
\href{http://arxiv.org/abs/gr-qc/9403028}{{\tt arXiv:gr-qc/9403028 [gr-qc]}}.

\bibitem{Jacobson:1993vj}
T.~Jacobson, G.~Kang, and R.~C. Myers, ``{On black hole entropy},''
  \href{http://dx.doi.org/10.1103/PhysRevD.49.6587}{{\em Phys. Rev.} {\bf D49}
  (1994)  6587--6598},
\href{http://arxiv.org/abs/gr-qc/9312023}{{\tt arXiv:gr-qc/9312023 [gr-qc]}}.

\bibitem{Lewkowycz:2013nqa}
A.~Lewkowycz and J.~Maldacena, ``{Generalized gravitational entropy},''
  \href{http://dx.doi.org/10.1007/JHEP08(2013)090}{{\em JHEP} {\bf 08} (2013)
  090},
\href{http://arxiv.org/abs/1304.4926}{{\tt arXiv:1304.4926 [hep-th]}}.

\bibitem{York:1972sj}
J.~W. York, Jr., ``{Role of conformal three geometry in the dynamics of
  gravitation},''
\href{http://dx.doi.org/10.1103/PhysRevLett.28.1082}{{\em Phys. Rev. Lett.}
  {\bf 28} (1972)  1082--1085}.

\bibitem{Dong:2013qoa}
X.~Dong, ``{Holographic Entanglement Entropy for General Higher Derivative
  Gravity},'' \href{http://dx.doi.org/10.1007/JHEP01(2014)044}{{\em JHEP} {\bf
  01} (2014)  044},
\href{http://arxiv.org/abs/1310.5713}{{\tt arXiv:1310.5713 [hep-th]}}.

\bibitem{Camps:2013zua}
J.~Camps, ``{Generalized entropy and higher derivative Gravity},''
  \href{http://dx.doi.org/10.1007/JHEP03(2014)070}{{\em JHEP} {\bf 03} (2014)
  070},
\href{http://arxiv.org/abs/1310.6659}{{\tt arXiv:1310.6659 [hep-th]}}.

\bibitem{Bhattacharyya:2013gra}
A.~Bhattacharyya, M.~Sharma, and A.~Sinha, ``{On generalized gravitational
  entropy, squashed cones and holography},''
  \href{http://dx.doi.org/10.1007/JHEP01(2014)021}{{\em JHEP} {\bf 01} (2014)
  021},
\href{http://arxiv.org/abs/1308.5748}{{\tt arXiv:1308.5748 [hep-th]}}.

\bibitem{Balasubramanian:2001nb}
V.~Balasubramanian, J.~de~Boer, and D.~Minic, ``{Mass, entropy and holography
  in asymptotically de Sitter spaces},''
  \href{http://dx.doi.org/10.1103/PhysRevD.65.123508}{{\em Phys. Rev.} {\bf
  D65} (2002)  123508},
\href{http://arxiv.org/abs/hep-th/0110108}{{\tt arXiv:hep-th/0110108
  [hep-th]}}.

\bibitem{GeorgiWeak}
H.~Georgi, {\em {Weak Interactions}}.
\newblock \url{http://www.people.fas.harvard.edu/~hgeorgi/weak.pdf}.

\bibitem{Manohar:1996cq}
A.~V. Manohar, ``{Effective field theories},''
  \href{http://dx.doi.org/10.1007/BFb0104294}{{\em Lect. Notes Phys.} {\bf 479}
  (1997)  311--362},
\href{http://arxiv.org/abs/hep-ph/9606222}{{\tt arXiv:hep-ph/9606222
  [hep-ph]}}.

\bibitem{Arkani-Hamed:2015bza}
N.~Arkani-Hamed and J.~Maldacena, ``{Cosmological Collider Physics},''
\href{http://arxiv.org/abs/1503.08043}{{\tt arXiv:1503.08043 [hep-th]}}.

\bibitem{Lee:2016vti}
H.~Lee, D.~Baumann, and G.~L. Pimentel, ``{Non-Gaussianity as a Particle
  Detector},''
\href{http://arxiv.org/abs/1607.03735}{{\tt arXiv:1607.03735 [hep-th]}}.

\bibitem{Stueckelberg:1957zz}
E.~C.~G. Stueckelberg, ``{Theory of the radiation of photons of small arbitrary
  mass},''
{\em Helv. Phys. Acta} {\bf 30} (1957)  209--215.

\bibitem{Hinterbichler:2011tt}
K.~Hinterbichler, ``{Theoretical Aspects of Massive Gravity},''
  \href{http://dx.doi.org/10.1103/RevModPhys.84.671}{{\em Rev.Mod.Phys.} {\bf
  84} (2012)  671--710},
\href{http://arxiv.org/abs/1105.3735}{{\tt arXiv:1105.3735 [hep-th]}}.

\bibitem{Avramidi:1986mj}
I.~G. Avramidi, {\em {Covariant methods for the calculation of the effective
  action in quantum field theory and investigation of higher derivative quantum
  gravity}}.
\newblock PhD thesis, Moscow State U., 1986.
\newblock
\href{http://arxiv.org/abs/hep-th/9510140}{{\tt arXiv:hep-th/9510140
  [hep-th]}}.
\newblock

\bibitem{Dvali:2001gx}
G.~R. Dvali, G.~Gabadadze, M.~Kolanovic, and F.~Nitti, ``{Scales of gravity},''
  \href{http://dx.doi.org/10.1103/PhysRevD.65.024031}{{\em Phys. Rev.} {\bf
  D65} (2002)  024031},
\href{http://arxiv.org/abs/hep-th/0106058}{{\tt arXiv:hep-th/0106058
  [hep-th]}}.

\bibitem{Schwinger:1951nm}
J.~S. Schwinger, ``{On gauge invariance and vacuum polarization},''
\href{http://dx.doi.org/10.1103/PhysRev.82.664}{{\em Phys. Rev.} {\bf 82}
  (1951)  664--679}.

\bibitem{Cohen:2008wz}
T.~D. Cohen and D.~A. McGady, ``{The Schwinger mechanism revisited},''
  \href{http://dx.doi.org/10.1103/PhysRevD.78.036008}{{\em Phys. Rev.} {\bf
  D78} (2008)  036008},
\href{http://arxiv.org/abs/0807.1117}{{\tt arXiv:0807.1117 [hep-ph]}}.

\bibitem{Bekenstein:1980jp}
J.~D. Bekenstein, ``{A Universal Upper Bound on the Entropy to Energy Ratio for
  Bounded Systems},''
\href{http://dx.doi.org/10.1103/PhysRevD.23.287}{{\em Phys. Rev.} {\bf D23}
  (1981)  287}.

\bibitem{Bousso:1999xy}
R.~Bousso, ``{A Covariant entropy conjecture},''
  \href{http://dx.doi.org/10.1088/1126-6708/1999/07/004}{{\em JHEP} {\bf 07}
  (1999)  004},
\href{http://arxiv.org/abs/hep-th/9905177}{{\tt arXiv:hep-th/9905177
  [hep-th]}}.

\bibitem{Bousso:2002ju}
R.~Bousso, ``{The Holographic principle},''
  \href{http://dx.doi.org/10.1103/RevModPhys.74.825}{{\em Rev. Mod. Phys.} {\bf
  74} (2002)  825--874},
\href{http://arxiv.org/abs/hep-th/0203101}{{\tt arXiv:hep-th/0203101
  [hep-th]}}.

\bibitem{Kabat:1995eq}
D.~N. Kabat, ``{Black hole entropy and entropy of entanglement},''
  \href{http://dx.doi.org/10.1016/0550-3213(95)00443-V}{{\em Nucl. Phys.} {\bf
  B453} (1995)  281--299},
\href{http://arxiv.org/abs/hep-th/9503016}{{\tt arXiv:hep-th/9503016
  [hep-th]}}.

\bibitem{Donnelly:2012st}
W.~Donnelly and A.~C. Wall, ``{Do gauge fields really contribute negatively to
  black hole entropy?},''
  \href{http://dx.doi.org/10.1103/PhysRevD.86.064042}{{\em Phys. Rev.} {\bf
  D86} (2012)  064042},
\href{http://arxiv.org/abs/1206.5831}{{\tt arXiv:1206.5831 [hep-th]}}.

\bibitem{Ostrogradsky:1850}
M.~Ostrogradsky {\em Mem. Ac. St. Petersbourg} {\bf VI 4} (1850)  385.

\bibitem{Woodard:2006nt}
R.~P. Woodard, ``{Avoiding dark energy with 1/r modifications of gravity},''
  \href{http://dx.doi.org/10.1007/978-3-540-71013-4_14}{{\em Lect. Notes Phys.}
  {\bf 720} (2007)  403--433},
\href{http://arxiv.org/abs/astro-ph/0601672}{{\tt arXiv:astro-ph/0601672
  [astro-ph]}}.

\bibitem{Simon:1990ic}
J.~Z. Simon, ``{Higher Derivative Lagrangians, Nonlocality, Problems and
  Solutions},''
\href{http://dx.doi.org/10.1103/PhysRevD.41.3720}{{\em Phys. Rev.} {\bf D41}
  (1990)  3720}.

\bibitem{Jaen:1986iz}
X.~Jaen, J.~Llosa, and A.~Molina, ``{A Reduction of order two for infinite
  order lagrangians},''
\href{http://dx.doi.org/10.1103/PhysRevD.34.2302}{{\em Phys. Rev.} {\bf D34}
  (1986)  2302}.

\bibitem{Burgess:2014lwa}
C.~P. Burgess and M.~Williams, ``{Who You Gonna Call? Runaway Ghosts, Higher
  Derivatives and Time-Dependence in EFTs},''
  \href{http://dx.doi.org/10.1007/JHEP08(2014)074}{{\em JHEP} {\bf 08} (2014)
  074},
\href{http://arxiv.org/abs/1404.2236}{{\tt arXiv:1404.2236 [gr-qc]}}.

\bibitem{Drummond:1979pp}
I.~Drummond and S.~Hathrell, ``{QED Vacuum Polarization in a Background
  Gravitational Field and Its Effect on the Velocity of Photons},''
\href{http://dx.doi.org/10.1103/PhysRevD.22.343}{{\em Phys.Rev.} {\bf D22}
  (1980)  343}.

\bibitem{Goon:2016une}
G.~Goon and K.~Hinterbichler, ``{Superluminality, Black Holes and Effective
  Field Theory},''
\href{http://arxiv.org/abs/1609.00723}{{\tt arXiv:1609.00723 [hep-th]}}.

\bibitem{Donoghue:1994dn}
J.~F. Donoghue, ``{General relativity as an effective field theory: The leading
  quantum corrections},''
  \href{http://dx.doi.org/10.1103/PhysRevD.50.3874}{{\em Phys. Rev.} {\bf D50}
  (1994)  3874--3888},
\href{http://arxiv.org/abs/gr-qc/9405057}{{\tt arXiv:gr-qc/9405057 [gr-qc]}}.

\bibitem{BjerrumBohr:2002ks}
N.~E.~J. Bjerrum-Bohr, J.~F. Donoghue, and B.~R. Holstein, ``{Quantum
  corrections to the Schwarzschild and Kerr metrics},''
  \href{http://dx.doi.org/10.1103/PhysRevD.68.084005,
  10.1103/PhysRevD.71.069904}{{\em Phys. Rev.} {\bf D68} (2003)  084005},
  \href{http://arxiv.org/abs/hep-th/0211071}{{\tt arXiv:hep-th/0211071
  [hep-th]}}.
[Erratum: Phys. Rev.D71,069904(2005)].

\bibitem{Almheiri:2012rt}
A.~Almheiri, D.~Marolf, J.~Polchinski, and J.~Sully, ``{Black Holes:
  Complementarity or Firewalls?},''
  \href{http://dx.doi.org/10.1007/JHEP02(2013)062}{{\em JHEP} {\bf 02} (2013)
  062},
\href{http://arxiv.org/abs/1207.3123}{{\tt arXiv:1207.3123 [hep-th]}}.

\bibitem{Bjerrum-Bohr:2016hpa}
N.~E.~J. Bjerrum-Bohr, J.~F. Donoghue, B.~R. Holstein, L.~Plante, and
  P.~Vanhove, ``{Light-like Scattering in Quantum Gravity},''
\href{http://arxiv.org/abs/1609.07477}{{\tt arXiv:1609.07477 [hep-th]}}.

\bibitem{Bellazzini:2015cra}
B.~Bellazzini, C.~Cheung, and G.~N. Remmen, ``{Quantum Gravity Constraints from
  Unitarity and Analyticity},''
  \href{http://dx.doi.org/10.1103/PhysRevD.93.064076}{{\em Phys. Rev.} {\bf
  D93} (2016) no.~6, 064076},
\href{http://arxiv.org/abs/1509.00851}{{\tt arXiv:1509.00851 [hep-th]}}.

\bibitem{Duff:1974ud}
M.~J. Duff, ``{Quantum corrections to the schwarzschild solution},''
\href{http://dx.doi.org/10.1103/PhysRevD.9.1837}{{\em Phys. Rev.} {\bf D9}
  (1974)  1837--1839}.

\bibitem{El-Menoufi:2015cqw}
B.~K. El-Menoufi, ``{Quantum gravity of Kerr-Schild spacetimes and the
  logarithmic correction to Schwarzschild black hole entropy},''
  \href{http://dx.doi.org/10.1007/JHEP05(2016)035}{{\em JHEP} {\bf 05} (2016)
  035},
\href{http://arxiv.org/abs/1511.08816}{{\tt arXiv:1511.08816 [hep-th]}}.

\bibitem{Donoghue:2015nba}
J.~F. Donoghue and B.~K. El-Menoufi, ``{Covariant non-local action for massless
  QED and the curvature expansion},''
  \href{http://dx.doi.org/10.1007/JHEP10(2015)044}{{\em JHEP} {\bf 10} (2015)
  044},
\href{http://arxiv.org/abs/1507.06321}{{\tt arXiv:1507.06321 [hep-th]}}.

\bibitem{Goon:2016ihr}
G.~Goon, K.~Hinterbichler, A.~Joyce, and M.~Trodden, ``{Aspects of Galileon
  Non-Renormalization},''
\href{http://arxiv.org/abs/1606.02295}{{\tt arXiv:1606.02295 [hep-th]}}.

\bibitem{Abbott:2016blz}
{\bf Virgo, LIGO Scientific} Collaboration, B.~P. Abbott {\em et al.},
  ``{Observation of Gravitational Waves from a Binary Black Hole Merger},''
  \href{http://dx.doi.org/10.1103/PhysRevLett.116.061102}{{\em Phys. Rev.
  Lett.} {\bf 116} (2016) no.~6, 061102},
\href{http://arxiv.org/abs/1602.03837}{{\tt arXiv:1602.03837 [gr-qc]}}.

\end{thebibliography}\endgroup

\end{document}